\documentclass[review]{elsarticle}

\usepackage{lineno,hyperref}
\modulolinenumbers[5]

\journal{Journal of \LaTeX\ Templates}

\usepackage{booktabs}

\usepackage[listings, skins, theorems]{tcolorbox} 
\usepackage{etoolbox} 
\usepackage{xcolor} 
\usepackage{enumerate}
\usepackage{mathrsfs}
\usepackage{tikz, xstring, calc, pgfmath}
\usepackage{ifthen}
\usepackage{amsmath}
\usepackage{amsthm}
\usepackage{amsfonts}
\usepackage{amssymb}
\usepackage{algorithm}
\usepackage[noend]{algpseudocode}
\usepackage{subcaption}
\usepackage{graphicx}
\usepackage{multirow}
\usepackage{multicol}
\usepackage{array}
\usepackage{mathtools}
\usepackage{wasysym}
\usepackage{xspace}
\usepackage{arydshln} 

\usepackage{pifont} 

\usepackage{tikz}
\usetikzlibrary{automata}
\usetikzlibrary[arrows,decorations.pathmorphing,backgrounds,positioning,fit,petri]
\tikzset{
  bend angle=30,
  ->,
  shorten >=1pt,
  node distance=2cm and 2cm,
  on grid,
  auto,
  initial where=above,
  initial text=,
  initial distance=0.6cm,
  inner sep=0.5mm,
  smallstate/.style={circle,draw},
  rootstate/.style={rectangle,draw},
  openstate/.style={},
  succstate/.style={font={$\surd$}},
  bendanglelarge/.style={bend angle=45},
  bendanglesmall/.style={bend angle=10},
  dim/.style={lightgray},
  diag/.style={red},
  bisim/.style={red,dashed,-,bend left},
  loop above/.style={in=110,out=70,loop,distance=0.5cm},
  loop left/.style={in=200,out=160,loop,swap,distance=0.5cm},
  loop right/.style={in=20,out=-20,loop,distance=0.5cm},
  loop below/.style={in=290,out=250,loop,distance=0.5cm}
}

\newcommand{\naturals}{\mathbb{N}}

\usepackage{bbold}
\let\altmathbb\mathbb
\newcommand{\one}{\altmathbb{1}}
\newcommand{\zero}{\altmathbb{0}}
\newcommand{\D}{\altmathbb{D}}

\newcommand{\wb}{\mathsf{WB}}
\renewcommand{\sb}{\mathsf{SB}}
\newcommand{\bas}{\mathit{bas}}
\newcommand{\wbbas}{\mathit{bas}_{\mathit{TS}}^W}
\newcommand{\sbbas}{\mathit{bas}_{\mathit{TS}}^S}

\newcommand{\target}{{A_t}}

\newcommand{\state}{{s}}
\newcommand{\St}{{S}}

\newcommand{\bn}{{\sf BN}}

\newcommand{\update}{{\xi}}

\newcommand{\x}{X}
\newcommand{\f}{F}

\newcommand{\ts}{{\mathit{TS}}}

\newcommand{\control}{{\sf C}}

\renewcommand{\path}{\rho}

\newcommand{\true}{\mathit{true}}
\newcommand{\false}{\mathit{false}}

\newcommand{\hd}{\mathit{hd}}
\newcommand{\reach}{\mathit{reach}}

\usepackage{calc}
\usepackage{enumitem}
\makeatletter
\newcounter{tmp@cnt}
\newcommand*\@labelpunc{)}
\newcommand*\multiItem[1][2]{%
    \refstepcounter{enumi}
    \setcounter{tmp@cnt}{\value{enumi}}
    \addtocounter{enumi}{#1-1}
    \item[\thetmp@cnt--\theenumi\@labelpunc]}
\makeatother

\tcbuselibrary{xparse,listings}
 \numberwithin{dummy}{section}

\newtheorem{observation}{Observation}
\newtheorem{definition}{Definition}
\newtheorem{example}{Example}

\newtheorem{corollary}{Corollary}

\newtcblisting{codeBox}[2][]{
    arc=0pt, outer arc=0pt,
    listing only, 
    title=#2,
    enlarge bottom by=0.1cm,
    enlarge top by=0.1cm,
    #1,
}

\definecolor{topicColor}{HTML}{330036}
\newtcolorbox{topicBox}[2][]
{
  colframe = topicColor!30,
  colback  = topicColor!5,
  coltitle = topicColor!10!black,  
  title    = {#2},
  enlarge bottom by=0.1cm,
  enlarge top by=0.1cm,
  #1,
}

\definecolor{noteColor}{HTML}{222228}
\newtcolorbox{noteBox}[1][]
{
  colframe = noteColor!45,
  colback  = noteColor!12,
  coltitle = noteColor!20!black,
  #1,
  enlarge bottom by=0.1cm,
  enlarge top by=0.1cm,
}

\definecolor{todoColor}{rgb}{0.0, 0.48, 0.65}
\newtcolorbox{todoBox}[2][]
{
 arc=0pt, outer arc=0pt,
  colframe = todoColor!65,
  colback  = todoColor!25,
  coltitle = todoColor!20!white,
  title    = {#2},
  enlarge bottom by=0.1cm,
  enlarge top by=0.1cm,
}

\newtcolorbox{equationBox}[1][]{
  colback=gray!1!white,
  enlarge bottom by=0.1cm,
  enlarge top by=0.1cm
}
\AtBeginEnvironment{equationBox}{\footnotesize}

\newcolumntype{C}[1]{>{\centering\arraybackslash}p{#1}}
\newcolumntype{L}[1]{>{\raggedright\arraybackslash}p{#1}}
\newcolumntype{R}[1]{>{\raggedleft\arraybackslash}p{#1}}

\begin{document}

\begin{frontmatter}

\title{Target Control of Asynchronous Boolean Networks}
\tnoteref{Target Control of Asynchronous Boolean Networks}

\author[1]{Cui Su}

\author[1,2]{Jun Pang\corref{mycorrespondingauthor}}
\cortext[mycorrespondingauthor]{Corresponding author}
\ead{jun.pang@uni.lu}

\address[1]{Interdisciplinary Centre for Security, Reliability and Trust, University of Luxembourg, Esch-sur-Alzette, Luxembourg} 
\address[2]{
Faculty of Science, Technology and Medicine, University of Luxembourg, Esch-sur-Alzette, Luxembourg}

\begin{abstract}
We study the target control of asynchronous Boolean networks, to identify efficacious interventions that can drive the dynamics of a given Boolean network from any initial state to the desired target attractor. Based on the application time, the control can be realised with three types of perturbations, including instantaneous, temporary and permanent perturbations. We develop efficient methods to compute the target control for a given target attractor with three types of perturbations. We compare our methods with the stable motif-based control on a variety of real-life biological networks to evaluate their performance. We show that our methods scale well for large Boolean networks and they are able to identify a rich set of solutions with a small number of perturbations. 
\end{abstract}


\end{frontmatter}


\section{Introduction}
\label{sec:intro-target-control}
Cell reprogramming has great potential for treating the most devastating diseases characterised by diseased cells or a deficiency of certain cells. 
It is capable of reprogramming any kind of abundant cells in the human body into the desired cells to restore functions of the diseased organs~\cite{SD16,GD19,GMS19}. 
Cell reprogramming opens up a novel field in cell and tissue engineering and regenerative medicine.

A major challenge of cell reprogramming lies in the identification of effective target proteins or genes, the manipulation of which can trigger desired changes. 
Lengthy time commitment and high cost hinder the efficiency of experimental approaches, which perform brute-force tests of tunable parameters and record corresponding results~\cite{L16}.
This strongly motivates us to turn to mathematical modelling of biological systems, which allows us to identify key genes or pathways that can trigger desired changes using computational methods.

Boolean network, first introduced by Kauffman~\cite{KS69}, is a well-established modelling framework for gene regulatory networks and their associated signalling pathways. 
It has apparent advantages compared to other modelling frameworks~\cite{Aku18}. 
Boolean network provides a qualitative description of biological systems and thus evades the parametrisation problem, which often occurs in quantitative models, such as models of ordinary differential equations (ODEs). 
In Boolean networks, molecular species, such as genes and transcription factors, are described as Boolean variables. 
Each variable is assigned with a Boolean function, which determines the evolution of the variable. 
Boolean functions characterise activation or inhibition regulations between molecular species. 
The dynamics of a Boolean network is assumed to evolve in discrete time steps, moving from one state to the next, under one of the updating schemes, such as {\it synchronous} or {\it asynchronous}. 
Under the synchronous scheme, all the nodes update their values simultaneously at each time step; while under the asynchronous scheme, only one node is randomly selected to update its value at each time step. 
We focus on the asynchronous updating scheme since it can capture the phenomenon that biological processes occur at different time scales. 
The steady-state behaviour of the dynamics is described as {\it attractors}, to one of which the system eventually settles down. 
Attractors are hypothesised to characterise cellular phenotypes~\cite{HS01}. 
Each attractor has a {\it weak basin} and a {\it strong basin}. 
The weak basin contains all the states that can reach this attractor, 
while the strong basin includes the states that can only reach this attractor and cannot reach any other attractors of the network. 
In the context of Boolean networks, cell reprogramming is interpreted as a control problem: modifying the parameters of a network to lead its dynamics from the source state towards the desired attractor.

Control theories have been employed to modulate the dynamics of complex networks in recent years. 
Due to the intrinsic non-linearity of biological systems, control methods designed for linear systems, 
such as structure-based control methods~\cite{LSB11,GLDB14,CGCK16}, are not applicable -- 
they can both overshoot and undershoot the number of control nodes for non-linear networks~\cite{GR16}. 
For nonlinear systems of ODEs, Fiedler {\it et al.}~\cite{ABGD13,BAGD13,ZYA17} proved that the control of a feedback vertex set is sufficient to control the entire network; 
and Cornelius {\it et al.}~\cite{CKM13} proposed a simulation-based method to predict instantaneous perturbations 
that can reprogram a cell from an undesired phenotype to a desired one.
However, further study is required to figure out if these two methods can be lifted to control Boolean networks. 
Several methods based on semi-tensor product (STP)~\cite{LCL17,ZLLC18,LZHY16,ZLKS19,WSZS19,CLW16,YYCJ19,ZKF13} have been proposed to solve different control problems for Boolean control networks (BCNs) under the synchronous updating scheme. 
For synchronous Boolean networks, Kim {\it et al.}~\cite{KSK13} developed a method to compute a small fraction of nodes, 
called `control kernels', that can be modulated to govern the dynamics of the network;  and
Moradi {\it el al.}~\cite{MGF19} developed an algorithm guided by forward dynamic programming to solve the control problem. 
However, all these methods are not directly applicable to asynchronous Boolean networks.

\begin{figure}[!ht]
  \centering
  \begin{subfigure}{0.45\textwidth}
  \centering
    \includegraphics[width=.7\textwidth]{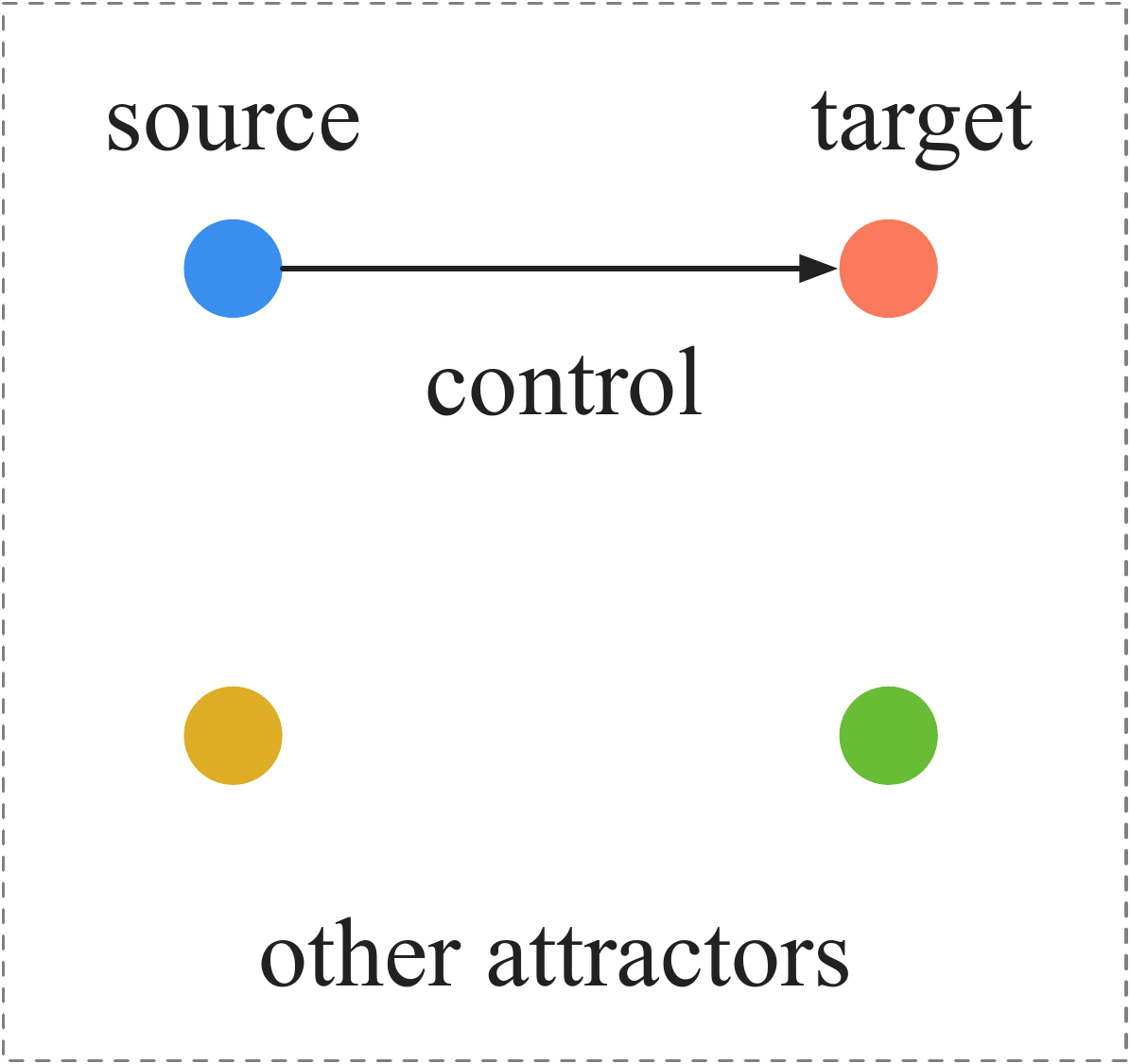} 
    \caption{Source-target control}
    \label{fig:stc-control}
  \end{subfigure}
  \begin{subfigure}{0.45\textwidth}
  \centering
    \includegraphics[width=.7\textwidth]{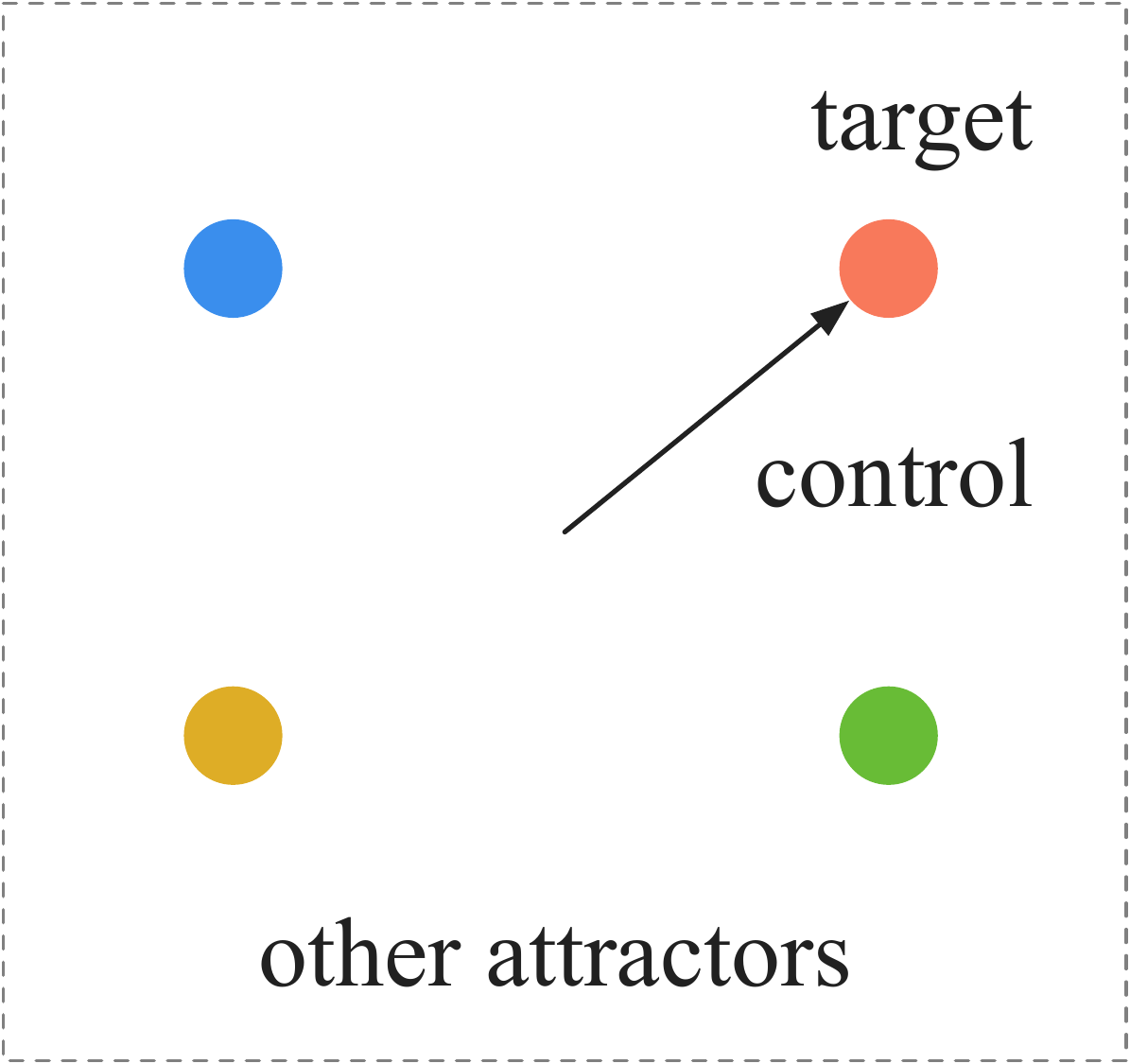}
    \caption{Target control}
    \label{fig:target}
  \end{subfigure}
  \caption[Source-target control and target control]%
  {(a) Source-target control and (b) target control of Boolean networks. }
  \label{fig:target-control}
\end{figure}

We have developed several methods~\cite{PSPM18,PSPM19,SPP19b,MSPPHP19,MSHPP19} for the {\it source-target control} of asynchronous Boolean networks: 
to drive the dynamics of a Boolean network from the source attractor to the target attractor. 
However, cells in tissues and in culture normally exist as a population of cells, 
corresponding to different states~\cite{SC14}. 
There is a need of {\it target control} to compute a subset of nodes, 
whose perturbation can drive the network from any initial state to the desired target attractor. 
Figures~\ref{fig:target-control} (a) and (b) illustrate the processes of source-target control and target control, respectively. 
The main difference lies in the source state: 
the source is a given attractor for source-target control, 
while the source can be any state in the state space for target control.

In this paper, we study target control of asynchronous Boolean networks with
instantaneous, temporary and permanent perturbations (ITC, TTC and PTC). 
We aim to find a control $\control=(\zero, \one)$,  
such that the instantaneous, temporary or permanent application of $\control$ -- setting the value of a node, 
whose index is in $\zero$ (or $\one$), to $0$ (or $1$) -- 
can drive the network from any initial state $s$ in the state space $S$ 
to the target attractor $\target$. 
Since the network can take any state $s\in S$ as the initial state, 
there exist a set of possible intermediate states with respect to $\control$ 
and they form a subset $S'$ of $S$, called {\it schema}. 
Instantaneous control should drive the system to states in the strong basin of the target attractor. 
Thus, we partition the strong basin of the target attractor into a set of disjoint schemata. 
The support variables of each schema form an instantaneous target control. 
For temporary and permanent control, due to their extended effects on the network dynamics, 
all the intermediate states should fall into the weak basin of the target attractor. 
Therefore, we partition the weak basin of the target attractor into a set of mutually disjoint schemata. 
Each schema results in a candidate temporary or permanent target control, which will be further optimised and verified.

Clinical applications are highly time-sensitive, controlling more nodes may shorten the period of time for generating sufficient desired cells for therapeutic application~\cite{GD19}. 
Hence, we integrate our method with a threshold $\zeta$ on the number of perturbations.  
By increasing $\zeta$, we can obtain solutions with at most $\zeta$ perturbations.   
It is worth noting that more perturbations may cause a significant increase in experimental costs, 
hence, 
the threshold $\zeta$ should be considered individually based on specific experimental settings.

Note that in our previous work~\cite{BCB20}, we have introduced the target control method with temporary perturbations, namely TTC. 
In this paper, which is an extended and revised version of~\cite{BCB20}, we further introduce the target control methods with instantaneous and permanent perturbations, i.e., ITC and PTC.
We implemented these three target control methods and compared their performance with 
the stable motif-based control (SMC)~\cite{ZA15} on various real-life biological networks.   
The results show that our methods outperform SMC in terms of the computational time for most of the networks. 
As for the temporary control, both our method TTC and SMC find a number of valid temporary controls, 
but our method is able to identify more solutions with fewer perturbations for some networks. 
Another interesting observation is that the number of required perturbations is often quite small compared to the sizes of the networks.  
This agrees with the empirical findings that the control of few nodes can reprogram biological networks~\cite{MS11}.

\section{Preliminaries}
\label{sec:preliminaries}
In this section, we present some preliminary notions of Boolean networks. 
\subsection{Boolean networks}
\label{ssec:bns}
A Boolean network (BN) describes elements of a dynamical system with binary-valued nodes and interactions between elements with Boolean functions. It is formally defined as:

\begin{definition}[Boolean networks]
A Boolean network is a tuple $\bn = (\x,\f)$ where $\x=\{x_1,x_2,\ldots, x_n\}$,
such that $x_i \in \x$ is a Boolean variable and $\f=\{f_1,f_2,\ldots,f_n\}$ is a set of Boolean functions over $\x$.
\end{definition}

The structure of a Boolean network $\bn = (\x,\f)$ can be viewed as a directed graph $G(V,E)$, 
called the {\it dependency graph} of $\bn$,
where $V=\{v_1,v_2\ldots, v_n\}$ is the set of {\it nodes}. 
Node $v_i \in V$ corresponds to variable $x_i \in \x$. 
For every $i,j\in \{1,2,\ldots,n\}$, there is a directed edge from $v_j$ to $v_i$, if and only if
$f_i$ depends on $x_j$. 
For the rest of the exposition, we assume an arbitrary but fixed network $\bn=(\x,\f)$ of $n$ variables is given to us.  
For all occurrences of $x_i$ and $f_i$, we assume $x_i$ and $f_i$ are elements of $\x$ and $\f$, respectively.

A {\it state} $\state$ of $\bn$ is an element in $\{0,1\}^n$.
Let $\St$ be the set of states of $\bn$. 
For any state $\state=(\state[1],\state[2],\ldots,\state[n])$, and for every $i \in \{1,2,\ldots,n\}$, the value of $\state[i]$,
represents the value that $x_i$ takes when the network is in state $\state$.
For some $i \in \{1,2,\ldots,n\}$, suppose $f_i$ depends on $x_{i_1},x_{i_2},\ldots, x_{i_k}$. 
Then $f_i(\state)$ will denote the value $f_i(\state[i_1],\state[i_2],\ldots, \state[i_k])$  
and $x_{i_1},x_{i_2},\ldots, x_{i_k}$ are called {\it parent nodes} of $x_i$, denoted as $par(x_i)$. 
For two states $\state,\state'\in \St$, the {\it Hamming distance} between $\state$ and $\state'$ will be denoted as $\hd(\state,\state')$
and $\arg(\hd(\state,\state'))\subseteq \{1,2,\ldots,n\}$ will denote the set of indices in which $\state$ and $\state'$ differ. 
For two subsets $ \St', \St''\subseteq \St$,
the Hamming distance between $\St'$ and $ \St''$ is defined as the minimum of the Hamming distances between all the states in $\St'$ and all the states in $\St''$.
That is, $\hd(\St', \St'')=\min_{\state' \in \St', \state'' \in \St''}\hd(\state',\state'')$.
We let $\arg(\hd(\St', \St''))$ denote the set of subsets of $\{1,2,\ldots,n\}$ such that $I\in \arg(\hd(\St', \St''))$
if and only if $I$ is a set of indices of the variables that realise this Hamming distance.

\begin{definition}[Control] \label{def:control}
A control $\control$ is a tuple $(\zero,\one)$, where $\zero, \one \subseteq \{1,2,\ldots,n\}$ and $\zero$ and $\one$ are mutually disjoint (possibly empty) sets of indices of nodes of a Boolean network $\bn$. 
The size of the control $\control$ is defined as $|\control|=|\zero|+|\one|$. 
Given a state $\state \in \St$, the application of $\control$ to $\state$, denoted as $\control(\state)$, 
is defined as a state $\state' \in \St$, such that $\state'[i]=0$ for $i \in \zero$ 
and $s'[i]=1$ for $i \in \one$. 
State $s'$ is called the intermediate state w.r.t. $\control$.
\end{definition}

The control can be lifted to a subset of states $\St' \subseteq \St$. 
Given a target control $\control=(\zero,\one)$, $\control(\St')=\St''$, where $\St''=\{s''\in \St|s''=\control(s'), s'\in \St'\}$. 
Set $\St''$ includes all the intermediate states with respect to $\control$. 
Intuitively, sets $\zero$ and $\one$ represent the indices of variables of $\bn$
whose values are held fixed to $0$ and $1$ respectively under the control $\control$. 
The application of a control $ \control $ to $\bn=(X,F)$ has the effect of reducing the state space of $\bn$ to those
which have the values of the variables
in $\zero$ and $\one$ set respectively to $0$ and $1$ and modifying the Boolean functions accordingly.
This results in a new Boolean network derived from $\bn$ defined as follows.

\begin{definition}[Boolean networks under control] \label{def:BNcontrol}
Let $\control=(\zero,\one)$ be a control and $\bn=(X,F)$ be a Boolean network. 
The Boolean network $\bn$ under control $\control$, denoted $\bn|_\control$, is defined as a tuple $\bn|_\control=(\hat{X},\hat{F})$, 
where $\hat{X}=\{\hat{x}_1,\hat{x}_2,\ldots, \hat{x}_n\}$ and $\hat{F}=\{\hat{f}_1,\hat{f}_2,\ldots,\hat{f}_n\}$, 
such that for all $i \in \{1,2,\ldots,n\}$: \\
(1) $\hat{x}_i=0$ if $i\in \zero$, $\hat{x}_i=1$ if $i\in \one$, and $\hat{x}_i=x_i$ otherwise; \\
(2) $\hat{f}_i=0$ if $i\in \zero$, $\hat{f}_i=1$ if $i\in \one$, and $\hat{f}_i=f_i$ otherwise.
\end{definition}

The state space of $\bn|_\control$, denoted $ \St|_\control$, is derived by fixing the values of the variables in $\control$ to their respective
values and is defined as $ \St|_\control=\{ s\in \St\ |\  s[i]=1 \text{ if } i\in \one \text{ and }  s[j]=0 \text{ if } j\in \zero\}$. 
It is obvious that $ \St|_\control\subseteq \St$.
For any subset $\St'$ of $\St$, we let $ \St'|_\control = \St'\cap \St|_\control$.

\subsection{Dynamics of Boolean networks}
\label{ssec:bns-dynamics}
In this section, 
we define several notions that can be interpreted on both $\bn$ and $\bn|_\control$. 
We use the generic notion $\bn=(\x,\f)$ to represent either $\bn=(\x,\f)$ or $\bn|_\control=(\hat{X},\hat{F})$. 
We assume that a Boolean network $\bn=(\x,\f)$ evolves in discrete time steps.
It starts from an initial state $s_0$ and its state changes in every time step 
based on the Boolean functions $\f$ and the updating schemes.
Different updating schemes lead to different dynamics of the network~\cite{MPSY17,ZH14}.
In this work, we are interested in the {\it asynchronous updating scheme} as it allows biological processes to happen at different classes of time scales and thus is more realistic.
We define {\it asynchronous dynamics} of Boolean networks as follows. 

\begin{definition}[Asynchronous dynamics of Boolean networks]\label{def:dynamics}
Suppose $\state_0\in \St$ is an initial state of $\bn$.
The asynchronous evolution of $\bn$ is a function $\update_\bn: \naturals \rightarrow \wp(\St)$
such that $\update_\bn(0)=\{\state_0\}$ and for every $j\geq 0$,
if $\state\in\update_\bn(j)$ then $\state'\in \update_\bn(j+1)$ is a possible {\em next state} of $\state$
iff either $\hd(\state,\state') = 1$ and there exists $i$ 
such that $\state'[i]=f_i(\state)=1-\state[i]$, 
or $\hd(\state,\state')=0$ and there exists $i$ such that $\state'[i]=f_i(\state)=s[i]$.
\end{definition}

It is worth noting that the asynchronous dynamics is non-deterministic.  
At each time step, only one node is randomly selected to update its value based on its Boolean function. 
A different choice may lead to a different next state $\state' \in \update(j+1)$. 
Henceforth, when we talk about the dynamics of a Boolean network, 
we shall explicitly mean the asynchronous dynamics. 
We describe the dynamics of a Boolean network as a {\it transition system (TS)}, defined as follows. 
\begin{definition}[Transition system of Boolean networks]\label{def:ts}
The transition system of a Boolean network $\bn$, denoted as $\ts$, is a tuple $(\St,\rightarrow_\bn)$,
where the vertices are the set of states $\St$ and for any two states $\state$ and $\state'$
there is a directed edge from $\state$ to $\state'$, denoted $\state \rightarrow \state'$, 
iff $\state'$ is a possible next state of $\state$ according to the asynchronous evolution function $\update$ of $\bn$.
\end{definition}

Similarly, we denote the transition system of a Boolean network under control, $\bn|_\control$, as $\ts|_\control$. 

\begin{figure}[!t]
\centering
\includegraphics[width=.9\textwidth]{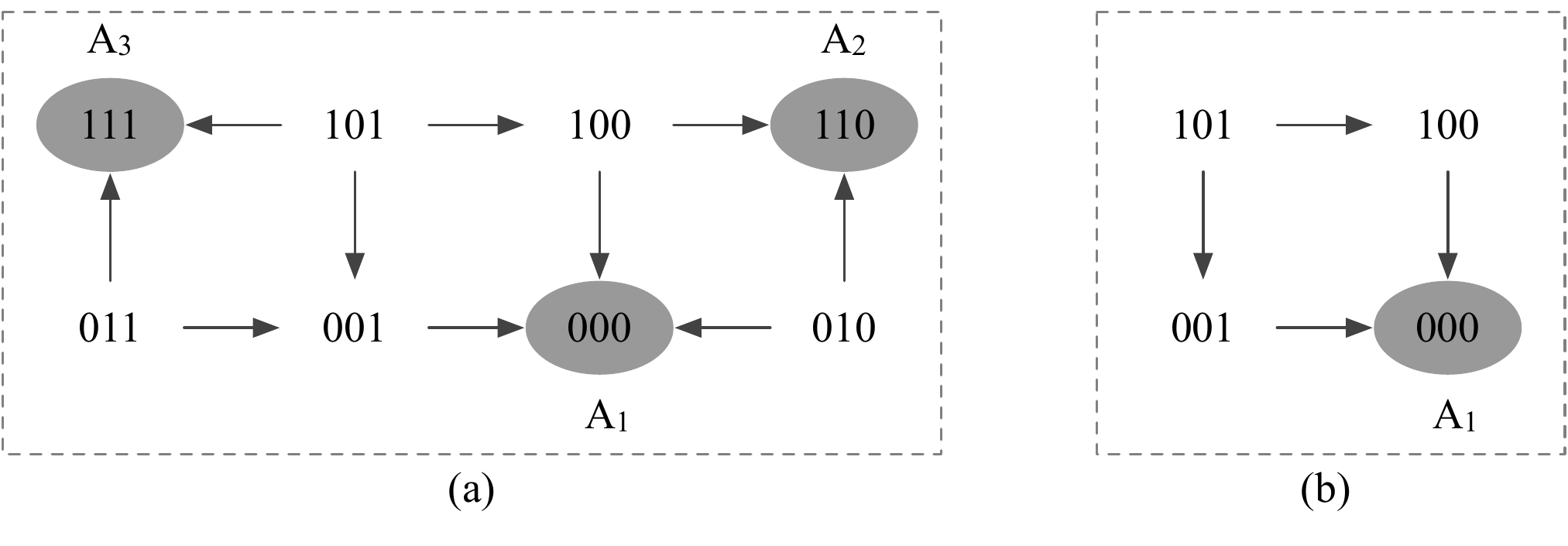}
\caption[The original transition system and the transition system under control of Example~\ref{example:bn} ]{(a) Transition system $\ts$ and 
(b) transition system under control $\ts|_\control$ of Example~\ref{example:bn}, where $\control=\{x_2=0\}$. 
We omit selfloops for all the states except for state $(101)$ in (a).}
\label{fig:example-control}
\end{figure}

\begin{example} \label{example:bn}
Consider a Boolean network $\bn=(\x,\f)$, where $\x=\{x_1,x_2,x_3\}$, $\f=\{f_1,f_2,f_3\}$,
and $f_1=x_2$,
$f_2=x_1$ and
$f_3=x_2 \land x_3$. 
The transition system $\ts$ of $\bn$ is given in Figure~\ref{fig:example-control}(a). 
Because the updating of the nodes is non-deterministic, 
a state can have more than one out-going edges. 

Given a control $\control=(\zero, \one)$, where $\zero=\{2\}, \one=\emptyset$ (i.e. $\{x_2=0\}$), 
the application of $\control$ reshapes the transition system $\ts$ of $\bn$ from Figure~\ref{fig:example-control}(a) 
to the transition system under control $\ts|_\control$ of $\bn|_\control$ in Figure~\ref{fig:example-control}(b). 
We can see that the control results in a new transition system, 
where only a subset of states and transitions are preserved. 
Therefore, the attractors of $\ts$ and $\ts|_\control$ might differ. 
For this example, only attractor $A_1$ is preserved in $\ts|_\control$ as shown in Figure~\ref{fig:example-control}(b). 
\end{example}

\subsection{Attractors and basins}
\label{ssec:attractor}
A {\em path} $\path$ from a state $ s$ to a state $ s'$ is a (possibly empty) sequence of transitions from $ s$ to $ s'$ in $\ts$, 
denoted $\path=s \rightarrow s_1 \rightarrow \ldots \rightarrow s'$.
A path from a state $ s$ to a subset $ S'$ of $ S$ is a path from $ s$ to any state $ s'\in  S'$. 
An {\em infinite path} $\path$ from $s$, $\path=s \rightarrow s_1 \rightarrow \ldots $, 
is a sequence of infinite transitions from $s$.   
A state $s' \in S $ appears {\it infinitely often} in $\path$ if for any $i \geq 0$, 
there exists $j \geq i$ such that $s_j=s'$. 
We assume every infinite path $\path$ is {\it fair} -- 
for any state $s'$ that appears infinitely often in $\path$, 
every possible next state $s''$ of $s'$ also appears infinitely often in $\path$. 
For a state $ s\in S$, $\reach(s)$ denotes the set of states $ s'$ such that there is a path 
from $s$ to $s'$ in $\ts$. 

\begin{definition}[Attractor]\label{def:attractor}
An attractor $A$ of $\ts$ (or of $\bn$) is a minimal non-empty subset of states of $ S$ such that for every state $ s\in A,~\reach(s)=A$.
\end{definition}

\begin{figure}
\centering
\includegraphics[width=0.90\textwidth]{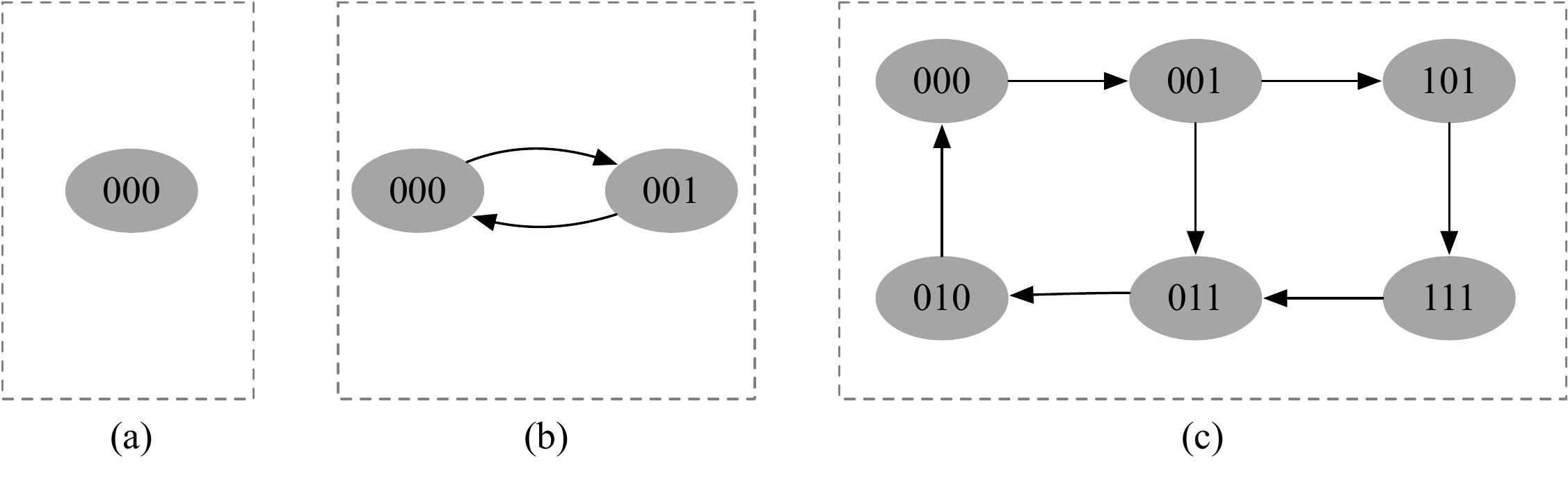}
\caption[Different types of attractors]{Different types of attractors of an asynchronous Boolean network: 
(a) a singleton attractor, (b) a simple loop, and (c) a complex loop. We omit selfloops for all the states.}
\label{fig:attractors}
\end{figure}

Attractors are hypothesised to characterise the steady-state behaviour of the network. 
Any state which is not part of an attractor is a transient state.
An attractor $A$ of $\ts$ is said to be reachable from a state $ s$ 
if $\reach(s)\cap A\neq\emptyset$.
The network starting at any initial state $s_0 \in S$ will eventually end up in one of the attractors of $\ts$ and remain there forever unless perturbed.
Under the asynchronous updating scheme, there are singleton attractors and cyclic attractors.  
Cyclic attractors can be further classified into:  
(1) a simple loop, in which all the states form a loop and every state appears only once per traversal through the loop; 
and (2) a complex loop, which has an intricate topology and includes several loops. 
Figures~\ref{fig:attractors} $(a)$, $(b)$ and $(c)$  
show a singleton attractor, a simple loop and a complex loop, respectively. 
Let $\mathcal{A}$ denote all the attractors of $\ts$. 
For an attractor $A,~A\in \mathcal{A}$, 
we define its {\it weak basin} as $\bas^W_\ts(A) = \{s\in S\ |\ \reach(s)\cap A\neq \emptyset\}$; 
the {\it strong basin} of $A$ is defined as $\bas^S_\ts(A) = \{s\in S\ |\ \reach(s)\cap A\neq \emptyset 
\text{ and } \reach(s) \cap A' = \emptyset  \text{ for any } A' \in \mathcal{A}, A'\neq A\}$. 
Intuitively, the weak basin of $A$, $\bas^W_\ts(A)$, 
contains all the states $s$ from which there exists at least one path to $A$,  
and there may also exist paths from $s$ to other attractors $A'~(A' \neq A)$ of $\ts$. 
The strong basin of $A$, $\bas^S_\ts(A)$, consists of all the states from which there only exist paths to $A$. 

\begin{example} \label{example:attractors}
The network in Example~\ref{example:bn} has three attractors $A_1=\{000\}$, $A_2=\{110\}$ and $A_3=\{111\}$, 
indicated as dark grey nodes in Figure~\ref{fig:example-control}(a). 
For attractor $A_1$, its strong basin, $\bas^S_\ts(A_1)$, contains two states $\{000, 001\}$; 
its weak basin, $\bas^W_\ts(A_1)$, contains six states $\{000, 001, 101, 011, 100, 010\}$.
\end{example}


\section{The target control problems}
\label{sec:target-control}
We have studied the source-target control of Boolean networks~\cite{PSPM18,PSPM19,SPP19b,MSPPHP19,MSHPP19,BCB20,SP2020},
to identify control paths that can drive the dynamics of the network 
from the source attractor to the target attractor. 
When the source is not given, 
to identify a subset of nodes, 
the control of which can stir the dynamics from any state $s \in \St$ to the target attractor $A_t$, 
is called {\it target control} of Boolean networks. 
Target control neglects the values of the nodes in $\state$, 
the application of a target control $\control$ inhibits the nodes in $\zero$ and overexpresses the nodes in $\one$. 

For target control, 
when perturbations are applied instantaneously, temporarily or permanently, 
we call it {\it instantaneous target control (ITC)}, {\it temporary target control (TTC)} or {\it permanent target control (PTC)}, respectively.   
Let $\bn$ be a given Boolean network, $\St$ be the set of states of $\bn$ and $\target$ be the target attractor of $\bn$. 
We formally define the three target control problems, ITC, TTC and PTC, as follows. 

\begin{definition}[Target control]\label{def:target-control}
\noindent
\begin{enumerate}
  \item {\bf Instantaneous target control (ITC)}:
  find a control $\control=(\zero,\one)$ such that 
  the dynamics of $\bn$ always eventually reaches $\target$
  on the instantaneous application of $\control$ to any initial state $\state, \state \in \St$.

  \item {\bf Temporary target control (TTC):}  
  find a control $\control=(\zero,\one)$ such that there exists a $t_0\geq 0$ such that for all $t\geq t_0$, 
  the dynamics of $\bn$ always eventually reaches $\target$ on the application of $\control$ to 
  any initial state $\state, \state \in \St$ for $t$ steps.

  \item {\bf Permanent target control (PTC):}  
  find a control $\control=(\zero,\one)$ such that the dynamics of $\bn$ always eventually reaches $\target$ 
  on the permanent application of $\control$ to any initial state $\state, \state \in \St$. (We assume implicitly that $\target$ is also an attractor of the transition system under control $\ts|_\control$). 

\end{enumerate}
\end{definition}

We define the concept of {\it schema}, which is crucial for the development of the target control methods. 
Given a control $\control=(\zero, \one)$, the possible intermediate states with respect to $\control$, denoted $\St'=\control(\St)$, 
form a schema, defined as follows.

\begin{definition}[Schema]
A subset $\St'$ of $\St$ is a schema if there exists a triple $M=(\zero, \one, \D)$, 
where $\zero \cup \one \cup \D = \{1,2,\ldots,n\}$, 
$\zero, \one$ and $\D$ are mutually disjoint (possibly empty) sets of indices of nodes of $\bn$, 
such that $\St'|_\zero=\{0\}^{|\zero|}$, $\St'|_\one=\{1\}^{|\one|}$ and $\St'|_\D=\{0,1\}^{|\D|}$. 
$\zero, \one$ and $\D$ are called off-set, on-set and don't-care-set of $\St'$, respectively.  
The elements in $\zero \cup \one$ are called indices of support variables of $\St'$. 
\end{definition}

Intuitively, for node $x_i, i \in \zero$, it has a value of $0$ in any state $s\in \St'$; 
for node $x_i, i \in \one$, it has a value of $1$ in any state $s\in \St'$. 
The projection of $\St'$ to the don't-care-set $\D$ contains all combinations of binary strings of $|\D|$ bits.
Thus, any schema $\St'$ is of size $2^{|\D|}$. 
Since the total number of nodes $n = |\zero| + |\one| + |\D|$ is fixed, 
a larger schema implies more elements in $\D$ and fewer elements in $\zero \cup \one$.

\begin{example}\label{example:tc}
Let us denote the values of the nodes in off-set, on-set and don't-care-set as `$0$', `$1$' and `$*$', respectively. 
For attractor $A_1$ of $\bn$ given in Example~\ref{example:bn}, 
its strong basin, $\sbbas(A_1)=\{000,001\}$, forms a schema, represented as `$00*$'. 
The weak basin, $\wbbas(A_1)=\{000, 001,010,011,101,100\}$, can be partitioned into two schemata
$\{000, 001, 010, 011\}$ and $\{101,100\}$, represented as `$0**$' and `$10*$', respectively.

\end{example}
\section{Instantaneous target control}
\label{sec:instantaneous-target-control}
An instantaneous control $\control$ will surely guide the dynamics of $\bn$ from any initial state $\state$ 
to the target attractor $\target$ if the intermediate state $\state'=\control(\state)$ is in the strong basin of $\target$ in $\ts$. 
Thus, when the initial state can be any state $\state \in \St$, 
to guarantee the inevitable reachability of the target attractor $\target$ on the instantaneous application of $\control$ to any $\state \in \St$, 
all possible intermediate states $\St'=\control(\St)$ must fall in 
the strong basin of the target attractor $\target$. 
Based on the theorem in~\cite{PSPM18}, we can derive the following corollary. 

\begin{corollary}\label{corollary:itc}
A control $\control=(\zero, \one)$ is an instantaneous target control from any initial state $\state \in \St$ 
to a target attractor $\target$ iff 
$\control(\St) \subseteq \sbbas(\target)$. 
\end{corollary}

Instantaneous control is only applied instantaneously, 
thus, its impact on the transition system is transient. 
If the instantaneous control does not drive the dynamics directly to $\sbbas(\target)$ 
but to any state $\state' \in (\St \setminus \sbbas(\target)) $, 
from $\state'$, there exist paths to some other attractor $A, A \neq \target$, based on the definition of strong basin. 
This does not ensure the inevitable reachability of the target attractor. 
Therefore, for an ITC $\control$, 
its intermediate states $\St'= \control(\St)$ must form a subset of $\sbbas(\target)$. 
For any possible intermediate state $\state' \in \St' $, $\state'[i]=0$ for $i \in \zero$ and $\state'[i]=1$ for $i \in \one$, 
which indicates that $\St'|_\zero=\{0\}^{|\zero|}$ and $\St'|_\one=\{1\}^{|\one|}$. 
Let $\D$ denote the indices of the nodes that are not in $\zero$ or $\one$. 
Then, $\St'|_\D=\{0,1\}^{|\D|}$ because the values of the nodes in the initial states $\St|_\D$ stay unchanged. 
In another word 
\begin{observation}\label{observation:schema}
If the initial state can be any state $\state \in \St$, 
for any control $\control=(\zero,\one)$, 
the set of intermediate states $\St'=\control(\St)$ forms a schema. 
\end{observation}

The notion of schema sheds light on the computation of ITC. 
Each schema $W_i$ of the strong basin of the target attractor, $\sbbas(\target)$, returns a candidate target control $\control_i=(\zero_i, \one_i)$, where $\zero_i$ and $\one_i$ are the off-set and on-set of $W_i$.  
The size of control $|\control_i|$ equals $(n-\log_2|W_i|)$, therefore, 
a larger schema results in a smaller control set. 
Thus, we can partition the strong basin of the target attractor, $\sbbas(\target)$, into a set of mutually disjoint schemata 
$\mathcal{W}=\{W_1, W_2, \ldots, W_m\}$, such that $W_1 \cup W_2 \cup \ldots \cup W_m = \sbbas(\target)$. 
Each $W_i \in \mathcal{W}$ is one of the largest schemata in $\sbbas(\target) \setminus (W_1 \cup \ldots \cup W_{i-1})$ and the indices of its support variables in $\zero_i$ and $\one_i$ form a candidate ITC $\control_i=(\zero_i,\one_i)$. 
In $\control_i$, the specified input nodes can be removed 
because input nodes do not have any predecessors and 
the values of the specified input nodes are fixed. 
For large networks, there may exist many valid control sets. 
To restrict the number and the size of solutions, 
we set a threshold $\zeta$ on the number of perturbations, 
keep $\zeta$ updated with the minimal size of the computed control sets, 
and only save the control sets with at most $\zeta$ perturbations.
Algorithm~\ref{alg:itc} realises the above idea in pseudocode. 

\begin{algorithm*}[!t]
\centering
\begin{algorithmic}[1]
\Procedure{{\sc Instantaneous\_Target\_Control}}{$\bn, \target$}
  \State initialise $\mathcal{L}:=\emptyset$ to store computed control sets 
  \State $I,I^s ,I^\mathit{ns} :=${\sc Comp\_input\_nodes}$(G)$ \hfill {\it // compute input nodes $I$, specified input nodes $I^s$ and non-specified input nodes $I^\mathit{ns}$}
  \State $\sb:=${\sc Comp\_Strong\_Basin}$(F,\target)$ \hfill{\it // the strong basin of $\target$ }
  \State $\mathcal{W}:=${\sc Comp\_schemata}$(\sb)$, $m:= |\mathcal{W}|$ 
  \State $\zeta:=n$ \hfill{\it an initial threshold on the number of perturbations} 

  \For {$i = 1: m$} \hfill {\it // traverse the set of schemata }
    \State $\control_i:=${\sc Comp\_support\_variables}$(W_i)$ \hfill{\it // $\control_i:=(\zero_i, \one_i)$}
      \State $\control_i := (\zero_i \setminus I^s, \one_i \setminus I^s)$ 
      \hfill{\it // remove specified input nodes}
      \If{$|\control_i| \leq \zeta$}
        \State save $\control_i$ to $\mathcal{L}$
        \State $\zeta:= \mathit{min}(|\control_i|, \zeta)$
      \EndIf
  \EndFor
  \State \Return $\mathcal{L}$
\EndProcedure
\end{algorithmic}
\caption{Instantaneous target control}
\label{alg:itc}
\end{algorithm*}

In this way, the computation of ITC is thus reduced to the computation of the strong basin of the target attractor and the computation of schemata. 
The computation of strong basins can be achieved efficiently with the procedure {\sc Comp\_Strong\_Basin}, which implements a decomposition-based approach
towards the computation of strong basins of Boolean networks (see~\cite{PSPM18,PSPM19} for details). 
The computation of schemata is based on BDDs, a symbolic representation of large state space.  
The size of a BDD is determined by both the set of states  
being represented and the chosen ordering of the variables.
In BDDs, a schema is represented as a {\it cube} and each state is the smallest cube, also called a {\it minterm}. 
To compute the largest schema $\St_i$ of $\St$ is equivalent to the computation of the largest cube of $\St$. 
The partitioning of the strong basin into schemata is then transformed into a cube cover problem in BDDs. 
A different variable ordering may lead to a different partitioning. 
Given a fixed ordering, the partitioning remains the same. 
Although finding the best variable ordering is NP-hard, 
there exist efficient heuristics to find the optimal ordering.
For our work, we compute a partitioning under one variable ordering as provided by the CUDD package~\cite{cudd} and we call this procedure {\sc Comp\_Schemata}.

\section{Temporary target control}
\label{sec:temporary-target-control}
In this section, we develop a method for TTC. 
First, we introduce the following corollary, which can be derived from the theorem in~\cite{SPP19b}.

\begin{corollary}\label{corollary:ttc}
A control $\control=(\zero, \one)$ is a temporary target control to a target attractor $\target$ from any initial state $s\in \St$ iff 
$\bas^\St_\ts(\target) \cap \St|_\control \neq \emptyset$ and $\control(\St) \subseteq \bas^\St_{\ts|_\control}(\bas^\St_\ts(\target) \cap \St|_\control)$.
\end{corollary}

Below, we give an intuitive explanation of Corollary~\ref{corollary:ttc}.  
We know that the application of a control $\control$ results in 
a new Boolean network $\bn|_\control$ and the state space is restricted to $\St|_\control$. 
To guarantee the inevitable reachability of $\target$, by the time we release the control, 
the network has to reach a state $s$ in the strong basin of $\target$ 
w.r.t. the original transition system $\ts$, i.e. $\bas^\St_\ts(\target)$, 
from which there only exist paths to $\target$. 
This requires the remaining strong basin in $\St|_\control$, i.e. $(\bas^\St_\ts(\target) \cap \St|_\control)$, is a non-empty set; 
otherwise, it is not guaranteed to reach $\target$. 
Furthermore, the condition $\control(\St) \subseteq \bas^\St_{\ts|_\control}(\bas^\St_\ts(\target) \cap \St|_\control)$ ensures that 
any possible intermediate state $s' \in \control(\St)$ is in the strong basin of the remaining strong basin $(\bas^\St_\ts(\target) \cap \St|_\control)$ in the transition system under control $\ts|_\control$, 
so that the network will always evolve to the remaining strong basin.  
Once the network reaches the remaining strong basin, 
the control can be released and the network will evolve spontaneously towards the target attractor $\target$. 
Based on the definition of the weak basin, 
it is sufficient to search the weak basin $\wbbas(\target)$ for TTC.

A noteworthy point is that temporary control needs to be released once the network reaches a state in $(\bas^\St_\ts(\target) \cap \St|_\control)$. 
On one hand, although Corollary~\ref{corollary:ttc} guarantees that partial of the strong basin of $\target$ in $\ts$ is preserved in $\ts|_\control$, it does not guarantee the presence of $\target$ in $\ts|_\control$. 
In that case, the control $\control$ has to be released at one point to recover the original $\ts$, which at the same time retrieves $\target$.  
On the other hand, in clinic, it is preferable to eliminate human interventions to avoid unforeseen consequences.
Concerning the timing to release the control, 
since it is hard to interpret theoretical time steps in diverse biological experiments, 
it would be more feasible for biologists to estimate the timing based on empirical knowledge and specific experimental settings.

Similar to ITC, the computation of TTC is also based on the concept of schema.  
Each schema $W_i$ of the weak basin $\wbbas(\target)$ gives a candidate TTC, $\control_i=(\zero_i, \one_i)$, for further optimisation and validation. 
A larger schema results in a smaller control set. 
To explore the entire weak basin $\wbbas(\target)$, 
we partition it into a set of mutually disjoint schemata 
$\mathcal{W}=\{W_1, W_2, \ldots, W_m\}$, $W_1 \cup W_2 \cup \ldots \cup W_m = \wbbas(\target)$. 
Each $W_i, i \in m$ is one of the largest schemata in $\wbbas(\target) \setminus (W_1 \cup \ldots \cup W_{i-1})$.
For $W_i$, the indices of its support variables in $\zero_i$ and $\one_i$ form a candidate control $\control_i=(\zero_i,\one_i)$.  
Each candidate control $\control_i$ is primarily optimised based on the properties of input nodes. 
Because input nodes do not have any predecessors,  
it is reasonable to assume that specified input nodes $I^s$ are redundant control nodes,  
while non-specified input nodes $I^\mathit{ns}$ are essential for control. 
For the remaining non-input nodes in $\control_i$, denoted $\control^r_i$, 
we verify its subsets of size $k$ based on Corollary~\ref{corollary:ttc} from $k=0$ with an increment of $1$, 
until we find a valid solution. 

\begin{algorithm*}[!ht]
\centering
\begin{algorithmic}[1]
\Procedure{{\sc Temporary\_Target\_Control}}{$\bn, \target$}
  \State initialise $\mathcal{L}:=\emptyset$ and $\Omega := \emptyset$ to store valid and checked control sets, resp.
  \State \label{line:input} $I, I^\mathit{ns} :=${\sc Comp\_input\_nodes}$(G)$ \hfill {\it //compute input nodes $I$ and non-specified input nodes $I^\mathit{ns}$.}
  \State \label{line:sb} $\sb:=${\sc Comp\_Strong\_Basin}$(F,\target)$ \hfill{\it //strong basin of $\target$ in $\ts$}
  \State\label{line:wb} $\wb:=${\sc Comp\_Weak\_Basin}$(F,\target)$ \hfill{\it //weak basin of $\target$ in $\ts$}
  \State \label{line:can}  $\mathcal{W}:=${\sc Comp\_schemata}$(\wb)$, $m:= |\mathcal{W}|$ 
  \State generate a vector $\Theta$ of length $m$ and set all the elements to $\false$
  \State $\zeta:=n$ \hfill{\it // a threshold on the number of perturbations. }

  \For {$i = 1: m$} \hfill {\it // traverse the set of schemata }
    \State \label{line:skip} {\sf if} {$\Theta[i]=\true$}, {\sf then} {\sf continue}
    \State \label{line:sup} $\control_i:=${\sc Comp\_support\_variables}$(W_i)$ \hfill{\it // $\control_i:=(\zero_i, \one_i)$}
      \State \label{line:ec} $\control^e_i := (\zero_i \cap I^\mathit{ns}, \one_i \cap I^\mathit{ns})$, 
      $\control^r_i := (\zero_i \setminus I, \one_i \setminus I)$ 
      \hfill{\it //essential control nodes and non-input nodes in $\control_i$}
      \State $k:=0$, $\mathit{isValid} := \false$
      \While{$\mathit{isValid} = \false$ and $k \leq \min(\zeta -|\control^e_i|, |\control^r_i|)$}
        \State $\mathcal{\control}^\mathit{sub}_i := ${\sc Comp\_subsets}$(\control^r_i,k)$  \hfill {\it //compute subsets of $\control^r_i$ of size $k$.}
        \For {$\control^\mathit{sub}_j \in \mathcal{\control}^\mathit{sub}_i$}
          \State $\control_i^j := \control^\mathit{sub}_j \cup \control^e_i$, $\Phi :=\control_i^j(\St) $ \hfill {\it // $\Phi$: intermediate states.}
          \If {$\control_i^j \notin \Omega$}  \hfill{\it //  $\control_i$ has not been checked.}
            \State $\mathit{isValid} := ${\sc Verify\_TTC}$(F, \control_i^j,\sb, \Phi)$
            \State add $\control_i^j$ to $\Omega$. 
            \If{$\mathit{isValid}=\true$}
              \State add $\control^j_i$ to $\mathcal{L}$, $\zeta := \min (\zeta, |\control^j_i|)$
              \State \label{line:remainW} $\Theta[z]:=\true$ if $W_z \subseteq \Phi$ for $z\in [i+1,m]$
              \hfill {\it // if a schema $W_z$ is a subset of $\Phi$, it will be skipped.}
            \EndIf
          \EndIf
        \EndFor
        \State {\sf if} $\mathit{isValid} = \false$, {\sf then} $k:=k+1$
      \EndWhile
  \EndFor
  \State \Return $\mathcal{L}$
\EndProcedure
\end{algorithmic}
\caption{Temporary target control}
\label{alg:ttc}
\end{algorithm*}

Algorithm~\ref{alg:ttc} implements the above idea in pseudocode. 
It takes as inputs the Boolean network $\bn=(\x, \f)$ and the target attractor $\target$.  
It first initialises two vectors $\mathcal{L}$ and $\Omega$ to store valid controls and the checked controls, respectively. 
(We use $\Omega$ to avoid duplicate control validations.)
Then, it computes input nodes $I$ and the non-specified input nodes $I^\mathit{ns},~ 
I^\mathit{ns} \subseteq I$ (line~\ref{line:input}). 
The weak basin $\wb$ and the strong basin $\sb$ of $\target$ of $\ts$ are computed using procedures
{\sc Comp$\_$Weak\_Basin} and {\sc Comp$\_$Strong\_Basin} developed in our previous work~\cite{PSPM18} (lines~\ref{line:sb}-\ref{line:wb}). 
The weak basin $\wb$ is then partitioned into $m$ mutually disjoint schemata 
with procedure {\sc Comp\_schemata}. 
Realisation of this procedure relies on the function 
to compute the largest cube provided by the CUDD package~\cite{cudd}. 
For each schema $W_i$, 
the indices of its support variables computed by procedure {\sc Comp\_support\_variables} 
form a candidate control $\control_i$ (line~\ref{line:sup}). 
The essential control nodes $\control^e_i$ of $\control_i$ consist of the non-specified input nodes 
and the non-input nodes in $\control_i$ constitute a set $\control^r_i$ for further optimisation (line $12$). 
We search the subsets of $\control^r_i$ starting from size $k=0$ with an increment of $1$ 
and verify whether the union of a subset $\control_j^\mathit{sub}$ of $\control_i^r$
and the essential nodes $\control^e_i$, namely $\control_i^j = \control_j^\mathit{sub} \cup \control_i^e$, 
is a valid temporary target control using procedure {\sc Verify\_TTC} in Algorithm~\ref{alg:verify}.
If $\control_i^j$ is valid, save it to $\mathcal{L}$. 
When all the subsets are traversed or a valid control has been found, 
we proceed to the next schema $W_{i+1}$. 
In the end, all the verified TTCs are returned. 

\begin{algorithm}[!t]
\centering
\begin{algorithmic}[1]
\Procedure{{\sc Verify\_TTC}}{$F, \control, \sb, \Phi$} \label{line:verify}
  \State $\mathit{isValid} := \false$
  \If {$\Phi \subseteq \sb$}
    \State $\mathit{isValid} = \true$
  \Else
    \State $\sb|_\control :=${\sc Comp\_state\_control}$(\control, \sb)$ \hfill {\it //compute the remaining strong basin w.r.t. $\control$ in $\ts|_\control$}
    \State $F|_\control := ${\sc Comp\_Fn\_control}$(\control, F)$ 
    \State $\bas^\St_{\ts|_\control}(\sb|_\control) := ${\sc Comp\_Strong\_Basin}$(F|_\control, \sb|_\control)$ 
    \If{$\Phi \subseteq \bas^\St_{\ts|_\control}(SB|_\control)$}
      \State $\mathit{isValid} = \true$
    \EndIf
  \EndIf
  \State \Return $\mathit{isValid}$
\EndProcedure
\end{algorithmic}
\caption{Verification of temporary target control}
\label{alg:verify}
\end{algorithm}

The most time-consuming part of our method lies in the verification process. 
As shown in Algorithm~\ref{alg:verify}, for each candidate control $\control$, 
we need to reconstruct the associated transition relations $F|_\control$ and compute the strong basin of the remaining strong basin in $\ts|_\control$, i.e. $\bas^\St_{\ts|_\control}(\sb|_\control)$ (lines $6$-$8$ of Algorithm~\ref{alg:verify}). 
Even though we have developed an efficient method for the strong basin computation, 
the computational time of Algorithm~\ref{alg:verify} still increases when the network size grows. 
To improve the efficiency, we propose two heuristics: 
(1) skip a schema $W_z$ (lines $10$ and $23$ of Algorithm~\ref{alg:ttc})
if it is a subset of intermediate states $\Phi$ of a pre-validated control $\control_i^j$ (line $23$ of Algorithm~\ref{alg:ttc});   
and (2) set a threshold $\zeta$ on the number of perturbations,  
keep $\zeta$ updated with the smallest size of valid TTCs $\control_i^j$ (line $22$ of Algorithm~\ref{alg:ttc}) 
and only compute control sets with at most $\zeta$ perturbations.

\section{Permanent target control}
\label{sec:permanent-target-control}
In this section, we develop a method to solve the problem of PTC. 
We first introduce the following corollary derived from the theorem in~\cite{SPP19b}. 

\begin{corollary}\label{corollary:ptc}
A control $\control=(\zero, \one)$ is a permanent target control from any initial state $s\in \St$ to a target attractor $\target$ iff 
$\target$ is an attractor of $\ts|_\control$ and 
$\control(\St) \subseteq \bas^\St_{\ts|_\control}(\target)$.
\end{corollary}

\begin{algorithm}[!t]
\centering
\begin{algorithmic}[1]
\Procedure{{\sc Verify\_PTC}}{$F, \control, \target, \Phi$} 
  \State $\mathit{isValid} := \false$
  \If {$\Phi|_\control = \target|_\control$} \label{line:preserveTarget}
    \State $F|_\control := ${\sc Comp\_Fn\_control}$(\control, F)$ 
    \State $\bas^\St_{\ts|_\control}(\target) := ${\sc Comp\_Strong\_Basin}$(F|_\control, \target)$ 
    \If{$\Phi \subseteq \bas^\St_{\ts|_\control}(\target)$}
      \State $\mathit{isValid} = \true$
    \EndIf
  \EndIf
  \State \Return $\mathit{isValid}$
\EndProcedure
\end{algorithmic}
\caption{Verification of permanent target control}
\label{alg:verify-ptc}
\end{algorithm}

Different from temporary control, permanent control is applied for all the following time steps. 
Thus, a permanent control should preserve the target attractor. 
To guarantee the inevitable reachability of the target attractor, 
all possible intermediate states should fall in the strong basin of the target attractor $\target$ 
in the transition system under control, $\ts|_\control$.

The algorithm for PTC can be derived from Algorithm~\ref{alg:ttc} by 
replacing procedure~{\sc Verify\_TTC} with procedure~{\sc Verify\_PTC} in Algorithm~\ref{alg:verify-ptc}. 
To avoid duplication, we only explain procedure~{\sc Verify\_PTC} here. 
This procedure is designed based on Corollary~\ref{corollary:ptc}. 
Line~\ref{line:preserveTarget} verifies whether the target attractor is preserved or not. 
If the target is preserved, we compute the transition relations under control $F|_\control$ 
and compute the strong basin of $\target$ in $\ts|_\control$ (lines 4-5). 
$\control$ is a PTC if the set of intermediate states is a subset of $\bas^\St_{\ts|_\control}(\target)$ (lines 6-7).

\begin{figure}[!t]
\centering
\includegraphics[width=.9\textwidth]{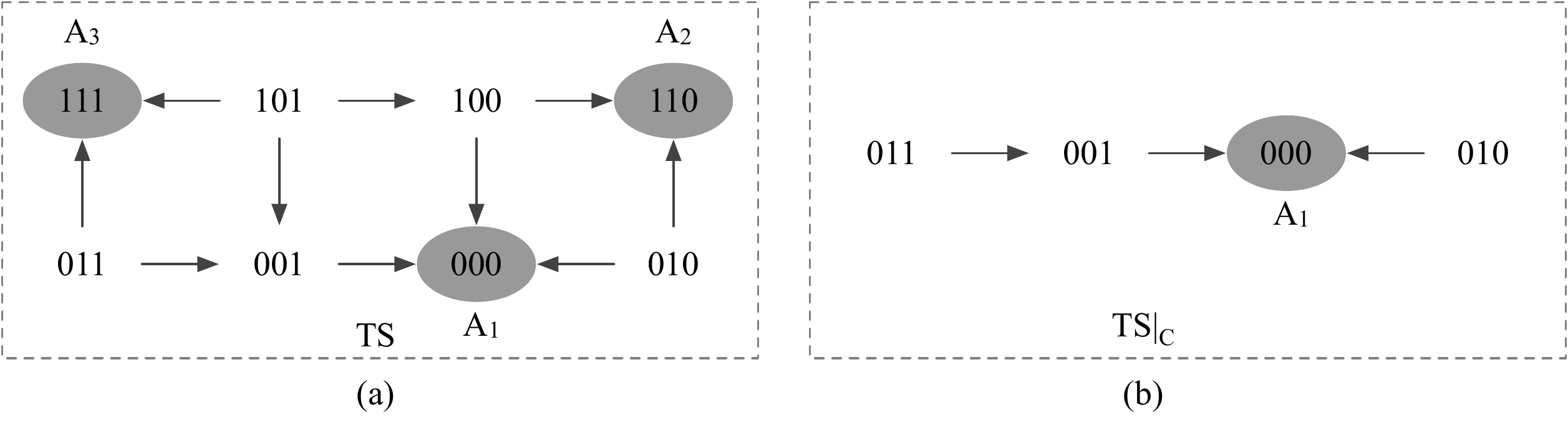}
\caption[Target control of Example~\ref{eg:tc}]{(a) Transition system $\ts$ and (b) transition system under control $\ts|_\control$ of Example~\ref{eg:tc}, where $\control=\{x_1=0\}$. 
We omit selfloops for all the states except for state $(101)$ in (a).}
\label{fig:example-tc}
\end{figure}

\begin{example}\label{eg:tc}
In Example~\ref{example:tc},  
we showed that the strong basin of $A_1$ of $\bn$, 
$\sbbas(A_1)$, can be represented as `$00*$'. 
It is easy to obtain the ITC for $A_1$, which is $\{x_1=0, x_2=0\}$. 
The simultaneous inhibition of $x_1$ and $x_2$ can drive the network 
from any state to $(000)$ or $(001)$, 
such that the network will eventually reach $(000)$. 

The weak basin of $A_1$, $\wbbas(A_1)$, can be divided into two schemata, represented as `$0**$' and `$10*$'. 
`$0**$' contains more states than `$10*$', which implies that `$0**$' can potentially give a smaller TTC or PTC. 
Algorithms for TTC and PTC verify subsets of the control derived from `$0**$' and `$10*$'. 
Based on Corollary~\ref{corollary:ttc} and Corollary~\ref{corollary:ptc}, $\{x_1=0\}$ is both a TTC and a PTC for $A_1$. 
By fixing $x_1$ to 0, the transition system changes from Figure~\ref{fig:example-tc} (a) to Figure~\ref{fig:example-tc} (b). 
The network is driven to a state in $\ts|_\control$ (see Figure~\ref{fig:example-tc} (b)) and will eventually stable in $A_1$. 
\end{example}

\section{Evaluation}
\label{sec:evaluation-target-control}
In this paper, we have developed three methods, ITC, TTC and PTC, 
for the target control of asynchronous Boolean networks. 
In particular, both TTC and the stable motif-based control (SMC) employ temporary perturbations. 
We apply our methods on a variety of biological networks and compare their performance with SMC. 
We discuss the results of the myeloid differentiation network and the cardiac gene regulatory network in detail 
and give an overview of the results of the other networks.
Our target control methods are implemented in our software CABEAN~\cite{cabean},
which contains a realisation of the decomposition-based detection method~\cite{MPQY17b,YMPQ19}
for identifying attractors in Boolean networks efficiently.\footnote{These attractor
detection methods were originally implemented in the software tool ASSA-PBN~\cite{assa,MPQY18,MPSY17}.} 
SMC\footnote{SMC is publicly available at \href{https://github.com/jgtz/StableMotifs}{https://github.com/jgtz/StableMotifs}.} is implemented in Java. 
All the experiments are performed on a high-performance computing (HPC) platform, which contains CPUs of Intel Xeon Gold 6132 @2.6 GHz.

\subsection{Control of the myeloid differentiation network}
\label{ssec:tc-myeloid}

\begin{table}[!t]
\centering
\begin{tabular}{L{0.8cm}C{2.2cm}C{2.2cm}C{2.2cm}C{2.2cm}}
\toprule
&   granulocytes    &    monocytes  &   megakaryocytes  &   erythrocytes \\ \midrule
ITC & 6  &  7 &  4  & 4 \\
TTC & 3  &  3 &  2  & 2 \\
PTC & 3  &  3 &  2  & 2\\
SMC & 4  &  4 &  2  & 2 \\
\bottomrule
\end{tabular}
\caption[The number of perturbations required by the target control of the myeloid differentiation network]{The number of perturbations required by ITC, TTC, PTC and SMC of the myeloid differentiation network.}
\label{tab:tc1_myeloid}
\end{table}

\begin{table}[!t]
\scalebox{0.85}{
\begin{tabular}{L{0.8cm}R{6cm}R{6cm}}
\toprule
  & granulocytes   & monocytes \\ \midrule
ITC &   \{GATA2=0 GATA1=0 Fli1=0 EgrNab=0 C/EBP$\alpha$=1 Gfi1=1\} &   
\{GATA2=0 GATA1=0 EgrNab=1 C/EBP$\alpha$=1 PU1=1 cJun=1 Gfi1=0\}  \\ \hline

\multirow{3}{*}{TTC} & \{C/EBP$\alpha$=1 PU1=1 cJun=0\} & \{EgrNab=1 C/EBP$\alpha$=1 PU1=1\} \\ 
& \{EgrNab=0 C/EBP$\alpha$=1 PU1=1\} & \{C/EBP$\alpha$=1 PU1=1 Gfi1=0\}  \\
& \{C/EBP$\alpha$=1 PU1=1 Gfi1=1\} & \\ \hline

\multirow{3}{*}{PTC} & \{C/EBP$\alpha$=1 PU1=1 cJun=0\} & \{EgrNab=1 C/EBP$\alpha$=1 PU1=1\} \\ 
& \{EgrNab=0 C/EBP$\alpha$=1 PU1=1\} &	\{C/EBP$\alpha$=1 PU1=1 Gfi1=0\}  \\
& \{C/EBP$\alpha$=1 PU1=1 Gfi1=1\}  & 
\\ \hline

\multirow{8}{*}{SMC} 
& \{GATA2=0 GATA1=0	&	\{GATA2=0 GATA1=0 \\
& C/EBP$\alpha$=1 EgrNab=0\}	&	C/EBP$\alpha$=1 EgrNab=1\} \\

& \{GATA2=0 GATA1=0	&	\{GATA2=0 GATA1=0 \\
& C/EBP$\alpha$=1 Gfi1=1\}	&	C/EBP$\alpha$=1 Gfi1=0\} \\

& \{GATA2=0 PU1=1 	&	\{GATA2=0 PU1=1 \\
& C/EBP$\alpha$=1 EgrNab=0\}	&	EgrNab=1 C/EBP$\alpha$=1\} \\

& \{GATA2=0 PU1=1	&	\{GATA2=0 PU1=1 \\ 
& C/EBP$\alpha$=1 Gfi1=1\}	&	Gfi1=0 C/EBP$\alpha$=1\} \\ 
\bottomrule
\end{tabular}}
\caption[Target control of granulocytes and monocytes]{The control sets computed by ITC, TTC, PTC and SMC for granulocytes and monocytes of the myeloid differentiation network.}
\label{tab:tc2_myeloid}
\end{table}

\begin{table}[!t]
\scalebox{0.85}{
\begin{tabular}{L{0.8cm}R{6cm}R{6cm}}
\toprule
 &  megakaryocytes  & erythrocytes \\ \midrule
ITC
&	\{GATA2=0 GATA1=0 EgrNab=1 C/EBP$\alpha$=1 PU1=1 cJun=1 Gfi1=0\} & 	\{GATA1=1 EKLF=1 Fli1=0 PU1=0\} \\  \hline

\multirow{5}{*}{TTC}
&	\{GATA2=1 EKLF=0\}	& 	\{GATA2=1 EKLF=1\} \\ 
&	\{GATA1=1 EKLF=0\}	& 	\{GATA1=1 EKLF=1\} \\ 
&	\{GATA2=1 Fli1=1\}	& 	\{GATA2=1 Fli1=0\} \\ 
&	\{GATA1=1 Fli1=1\}	& 	\{GATA1=1 Fli1=0\} \\ 
&	\{Fli1=1 PU1=0 \}	&	\\ \hline

\multirow{3}{*}{PTC}
&	\{GATA1=1 EKLF=0 \}	&	\{GATA1=1 EKLF=1 \} \\
&	\{GATA1=1 Fli1=1\}	& 	\{GATA1=1 Fli1=0 \} \\
&	\{Fli1=1 PU1=0 \}	&	\\ \hline

\multirow{2}{*}{SMC}	
&	\{GATA1=1 EKLF=0\}	&	\{GATA1=1 EKLF=1\} \\
&	\{GATA1=1 Fli1=1\}	&	\{GATA1=1 Fli1=0\} \\ \bottomrule
\end{tabular}}
\caption[Target control of megakaryocytes and erythrocytes]{The control sets computed by ITC, TTC, PTC and SMC for megakaryocytes and erythrocytes  of the myeloid differentiation network.}
\label{tab:tc3_myeloid}
\end{table}

The myeloid differentiation network is designed to model myeloid differentiation from common myeloid progenitors
to granulocytes, monocytes, megakaryocytes and erythrocytes~\cite{KMST11}. 
This network has $11$ nodes and $6$ attractors,  
4 of which correspond to granulocytes, monocytes, megakaryocytes and erythrocytes. 
We apply ITC, TTC, PTC and SMC to identify interventions to reach the four cell types. 

Table~\ref{tab:tc1_myeloid} gives the number of perturbations required by the four methods. 
The control with instantaneous perturbations (ITC) requires more perturbations than the control with temporary perturbations (TTC and SMC) and permanent perturbations (PTC) as expected. 
For granulocytes and monocytes, TTC and PTC find smaller control sets than SMC. 
For megakaryocytes and erythrocytes, 
TTC, PTC and SMC require the same number of perturbations.

Table~\ref{tab:tc2_myeloid} and Table~\ref{tab:tc3_myeloid} summarise the control sets identified by the four methods. 
For each attractor, ITC only finds one control set with more perturbations than the other methods. 
Although SMC finds more control sets than TTC for granulocytes and monocytes as shown in Table~\ref{tab:tc2_myeloid}, 
SMC requires four perturbations while TTC and PTC need only three perturbations. 
Since our methods TTC and PTC only compute the results within the threshold, 
they may identify more solutions if we increase the threshold to four. 
For megakaryocytes and erythrocytes in Table~\ref{tab:tc3_myeloid}, 
TTC, PTC and SMC require the same number of perturbations, 
but TTC provides more control sets than PTC and SMC. 
For this network, the results of PTC are either identical to TTC or just a subset of the solutions identified by TTC. 
Potentially, TTC is able to find smaller control sets than PTC, because it does not need to preserve the target attractor during the control. 
We will demonstrate this point in Section~\ref{ssec:tc-other_networks}. 

The total execution time of ITC, TTC, PTC and SMC for computing target control of this network are 0.003, 0.026, 0.03 and 8.178 seconds, respectively. 
We can see that our methods outperform SMC in efficiency.

\subsection{Control of the cardiac gene regulatory network}
\label{ssec:tc-cardiac}
The cardiac gene regulatory network integrates major genes that play essential roles 
in early cardiac development and FHF/SHF determination~\cite{HGZKK12}. 
It has $15$ nodes. 
This network consists of three attractors, two of which correspond to FHF and SHF, 
when the input node, exogenBMP2I, is set to 1~\cite{HGZKK12}. 
We apply ITC, TTC, PTC and SMC to identify interventions that can lead the network to FHF and SHF. 
The results of the four control methods are given in Table~\ref{tab:tc-cardiac}. 

\begin{table}[!t]
\centering
\scalebox{0.9}{
\begin{tabular}{L{0.8cm}R{4cm}R{7cm}}
\toprule 
  & SHF & FHF  \\ \midrule
ITC  & \{exogenCanWntI=1\} & \{canWnt=0 Foxc12=0 Tbx1=0 GATAs=0 Tbx5=1 exogenCanWntI=0 exogenCanWntII=0 \} \\ \hline
\multirow{6}{*}{TTC}  & \multirow{3}{*}{\{exogenCanWntI=1\}} 
& \{GATAs=1 exogenCanWntI=0 \} \\ 
&&\{Tbx5=1 exogenCanWntI=0\} \\
&&\{Nkx25=1 exogenCanWntI=0\} \\
&&\{Mesp1=1 exogenCanWntI=0\} \\
&&\{Tbx1=1 exogenCanWntI=0\} \\
&&\{Foxc12=1 exogenCanWntI=0\} \\\hline
\multirow{3}{*}{PTC}  & \multirow{3}{*}{\{exogenCanWntI=1\}} 
& \{GATAs=1 exogenCanWntI=0\} \\ 
&& \{Nkx25=1 exogenCanWntI=0\} \\ 
&& \{Tbx5=1 exogenCanWntI=0\} \\ \hline

\multirow{3}{*}{SMC}  & \multirow{3}{*}{\{exogenCanWntI=1\}} 
& \{GATAs=1 exogenCanWntI=0 \} \\ 
&&\{Tbx5=1 exogenCanWntI=0\} \\
&&\{Nkx25=1 exogenCanWntI=0\} \\ \bottomrule
\end{tabular}}
\caption[Target control of the cardiac network]{
The control sets computed by ITC, TTC, PTC and SMC for SHF and FHF of the cardiac gene regulatory network. }
\label{tab:tc-cardiac}
\end{table} 

With instantaneous, temporary or permanent perturbations, 
it is guaranteed to reach SHF by the control of the non-initialised input node, exogenCanWntI. 
To reach FHF, ITC requires seven perturbations,
while TTC realises the goal by the control of two nodes, 
including the input node exogenCanWntI together with one of the nodes in \{GATAs, Tbx5, Nkx25, Mesp1, Tbx1, Foxc12\}. 
The results of PTC and SMC are subsets of TTC, which demonstrates the ability of TTC in identifying more novel solutions. 
With temporary and permanent perturbations, the number of perturbations required to reach FHF is reduced from seven to two, 
which can greatly reduce experimental costs and improve the operability of the experiments.

The total execution time of ITC, TTC, PTC and SMC for computing target control for the three attractors 
are 0.035, 0.128, 0.148 and 4.540 seconds, respectively. 
For this network, our methods are more efficient than SMC.

\subsection{Control of other biological networks}
\label{ssec:tc-other_networks}
We apply the four target control methods, ITC, TTC, PTC and SMC, to some other networks introduced below to evaluate their performance. 
We give an overview of the number of nodes, the number of edges and the number of singleton and cyclic attractors of each network in Table~\ref{tab:overview-networks}. 
Details on the networks, such as the Boolean functions, are referred to the original works.

\begin{table}[!t]
\centering
\scalebox{1}{
\begin{tabular}{L{2.2cm}R{1cm}R{1cm} R{1.7cm}R{1.7cm}}
\toprule
\multirow{2}{*}{Network} & \multicolumn{1}{c}{\#} & \multicolumn{1}{c}{\#} 
& \multicolumn{1}{c}{\# singleton} & \multicolumn{1}{c}{\# cyclic}   \\ 
&nodes & edges  & attractors & attractors \\ \midrule
yeast & 10  & 28   & 12  & 1 \\ 
ERBB  & 20  & 52   & 3 & 0 \\ 
HSPC-MSC  & 26  & 81 & 2 & 2 \\ 
tumour  & 32  & 158  & 9 & 0 \\ 
{\small hematopoiesis}  & 33  & 88   & 5 & 0 \\ 
PC12  & 33  & 62   & 7 & 0 \\ 
bladder & 35  & 116  & 3 & 1 \\ 
PSC-bFA & 36  & 237  & 4 & 0 \\ 
{\small co-infection} & 52  & 136 & 30  & 0 \\ 
MAPK  & 53  & 105  & 12  & 0 \\ 
CREB  & 64  & 159  & 8 & 0 \\ 
HGF & 66  & 103  & 10  & 0 \\ 
{\small bortezomib} & 67  & 135  & 5 & 0 \\ 
T-diff  & 68  & 175  & 12  & 0 \\ 
HIV1  & 136 & 327  & 8 & 0 \\ 
CD4+  & 188 & 380  & 6 & 0 \\ 
pathway & 321 & 381 & 3 & 1 \\
\bottomrule
\end{tabular}}
\caption[An overview of the biological networks]{An overview of the biological networks.}
\label{tab:overview-networks}
\end{table}

\begin{itemize}
\item The cell cycle network of the fission yeast is constructed based on known biochemical interactions 
to recap regulations of the cell cycle of the fission yeast~\cite{DB08}. 

\item The ERBB receptor-regulated G1/S transition protein network combines ERBB signalling with G1/S transition of the mammalian cell cycle to identify new targets for breast cancer treatment~\cite{SFLK09}.

\item The HSPC-MSC network of $26$ nodes describes intercommunication pathways 
between hematopoietic stem and progenitor cells (HSPCs) 
and mesenchymal stromal cells (MSCs) in bone marrow (BM)~\cite{EMMP16}. 

\item The tumour network is built to study the role of individual mutations or their combinations in the metastatic process~\cite{CMC15}. 

\item The network of hematopoietic cell specification covers major transcription factors and signalling pathways for the development of lymphoid and myeloid \cite{COOAD17}.

\item The PC12 cell differentiation network~\cite{OKB16} 
  is a comprehensive model for the clarification of the cellular decisions towards proliferation or differentiation.
  It models the temporal sequence of protein signalling, transcriptional responses as well as the subsequent autocrine feedbacks~\cite{OKB16}.

\item The bladder cancer network of $35$ nodes allows one to identify the deregulated pathways and their influence on bladder tumourigenesis~\cite{WSA12}.

\item The model of mouse embryonic stem cells captures the signal-dependent emergence of cell subpopulations under different initial conditions~\cite{YOOLP18}.

\item The model of immune responses is constructed to study the immune responses against single and co-infections with the respiratory bacterium and the gastrointestinal helminth~\cite{TPMA12}.   

\item The MAPK network is constructed to study the MAPK responses to different stimuli and their contributions to cell fates~\cite{GCT13}. 

\item The CREB network is a complex neuronal network, whose output node is the transcription factor adenosine 3', 5'-monophosphate (cAMP) response element–binding protein (CREB)~\cite{ATE08}. 

\item The model for HGF-induced keratinocyte migration captures the onset and maintenance of hepatocyte growth factor-induced migration of primary human keratinocytes~\cite{SNB12}. 

\item The model of bortezomib responses can predict responses to the lower bortezomib exposure and the dose-response curve for bortezomib~\cite{COAM15}. 

\item The Th-cell differentiation network models regulatory elements and signalling pathways controlling Th-cell differentiation~\cite{NCCT10}.

\item The HIV-1 network models dynamic interactions between human immunodeficiency virus type $1$ (HIV-1) proteins 
and human signal-transduction pathways that are essential for activation of CD$4+$ T lymphocytes~\cite{ODRS14}. 

\item The CD$4^+$ T-cell network allows us to study downstream effects of CAV$1^{+/+}$, CAV$1^{+/-}$
and CAV$1^{-/-}$ on cell signalling and intracellular networks~\cite{conroy2014design}.

\item The model of signalling pathways central to macrophage activation integrates four crucial signalling pathways that are triggered early-on in the innate immune response~\cite{RRLP08}. 

\end{itemize}

\begin{table}[!t]
\centering
\begin{tabular}{L{2.2cm}R{1.5cm}R{1.5cm}R{1.5cm}R{1.5cm}}
\toprule
\multirow{2}{*}{Network} & \multicolumn{4}{c}{The minimal number of perturbations} \\ \cline{2-5}
 & ITC & TTC &  PTC &	SMC \\ \midrule
yeast	&	10	&	5	&	5	&	5 \\ 
ERBB	&	10	&	2	&	2	&	2 \\ 
HSPC-MSC	&	2	&	2	&	2	&	2 \\ 
{\small hematopoiesis}	&	5	&	3	&	3	&	$*$ \\ 
PC12	&	12	&	3	&	3	&	3 \\ 
bladder	&	14	&	2	&	2	&	4 \\ 
PSC-bFA	&	11	&	1	&	2	&	$*$ \\ 
{\small co-infection}	&	19	&	5	&	5	&	7 \\ 
MAPK	&	24	&	4	&	4	&	5 \\ 
CREB	&	3	&	3	&	3	&	$*$ \\ 
HGF	&	22	&	4	&	4	&	$*$ \\ 
{\small bortezomib}	&	3	&	1	&	1	&	$*$ \\ 
T-diff	&	20	&	4	&	4	&	4 \\ 
HIV1	&	3	&	3	&	3	&	$*$ \\ 
CD4+	&	7	&	3	&	3	&	3 \\ 
pathway	&	2	&	2	&	2	&	2 \\ 
\bottomrule
\end{tabular}
\caption[Target control of the biological networks]{The minimal number of perturbations required by ITC, TTC, PTC and SMC for several biological networks. }
\label{tab:tc1-networks}
\end{table}

\smallskip
\noindent
\textbf{Efficacy.}  
Table~\ref{tab:tc1-networks} summarises the minimal number of perturbations required by the four methods for one of the attractors of the networks. 
It is easy to observe that ITC requires more perturbations than TTC, PTC and SMC due to its instantaneous effect. 
ITC usually needs to control 10 to 20 nodes, 
whereas TTC, PTC and SMC can achieve the inevitable reachability of the target attractor with at most 7 perturbations. 
Moreover, it is hard to realise the simultaneous and instantaneous perturbation of a number of nodes, 
which makes the ITC less practical in application. 
Thus, TTC, PTC and SMC, which employ temporary or permanent perturbations, are preferable than ITC. 
For the bladder cancer network and the MAPK network, 
TTC and PTC identify smaller control sets than SMC. 
Compared to PTC, TTC has the ability to further reduce the number of perturbations as demonstrated by the model of mouse embryonic stem cells (PSC-bFA) -- the number of perturbations required by TTC and PTC are 1 and 2, respectively.

\begin{figure}[!t]
\centering
\includegraphics[width=\textwidth]{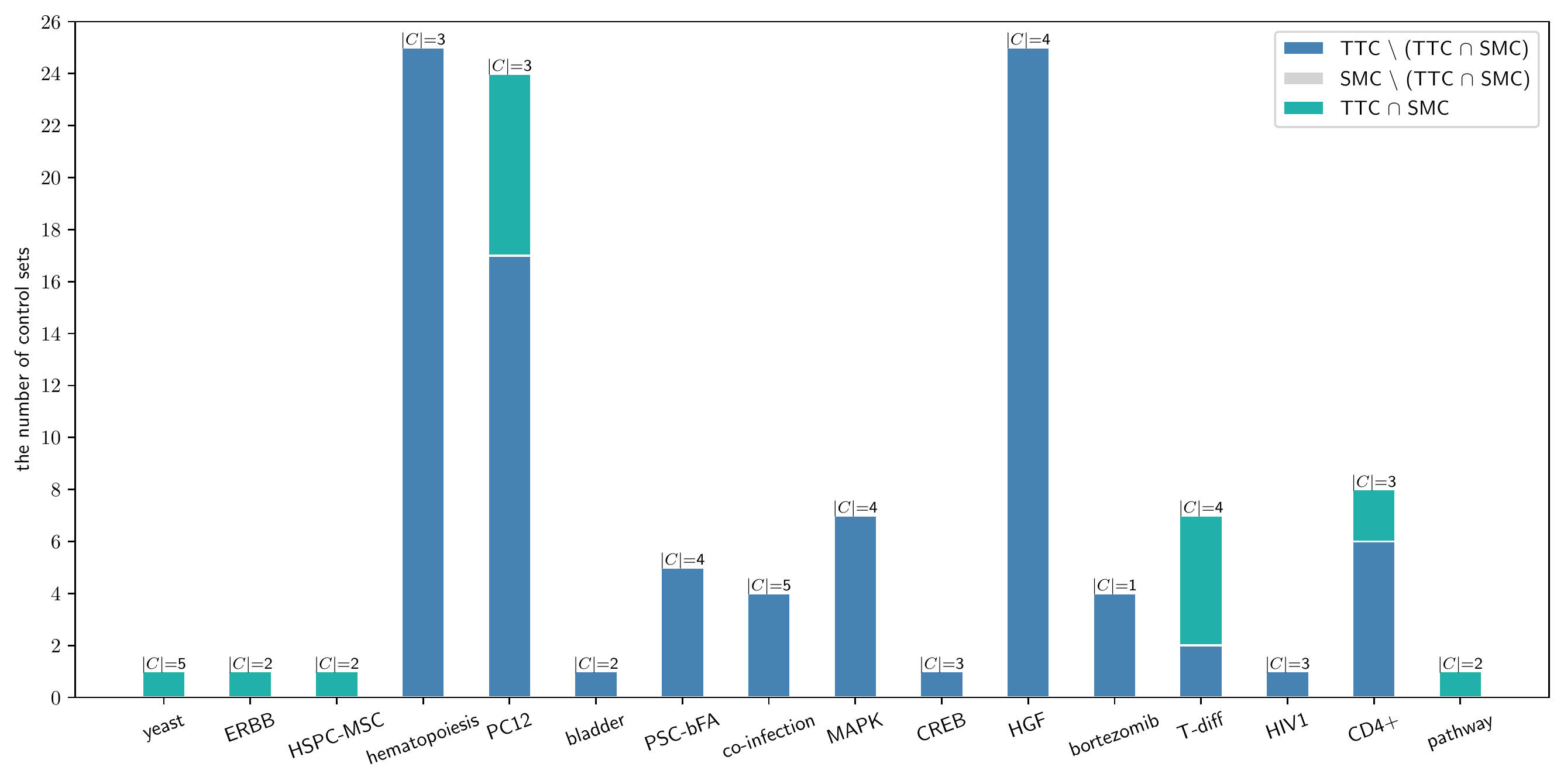}
\caption[Comparison of TTC and SMC on the number of solutions]{Comparison of TTC and SMC on the number of solutions. }
\label{fig:tc-networks}
\end{figure}

Both TTC and SMC solve the target control problem with temporary perturbations. 
To further compare these two methods, 
Figure~\ref{fig:tc-networks} shows the number of solutions identified by the two methods. 
The x-axis lists the names of the networks and the y-axis denotes the number of control sets. 
Blue bars and grey bars represent the control sets that only appear in the results of TTC and SMC 
(TTC$\setminus$ (TTC $\cap$ SMC), SMC$\setminus$ (TTC $\cap$ SMC)), respectively. 
Green bars represent the intersections of the two methods. 
Equations above the bars ($|\control|=k$) denote the number of nodes contained in the control set, 
i.e. the number of required perturbations.  

Since neither of the methods guarantees the minimal control, 
they may find control sets of different sizes for one attractor. 
For comparison, we only consider the smallest control sets. 
In Figure~\ref{fig:tc-networks}, there is no grey bar because 
the solutions identified by SMC are either also found by TTC and thus belong to (TTC $\cap$ SMC), 
or require more perturbations than TTC. 
For the cell cycle network of fission yeast (yeast), 
the ERBB receptor-regulated G1/S transition protein network (ERBB), 
the HSPC-MSC network (HSPC-MSC) and the model of signalling pathways (pathway), 
there are only green bars, which means that the results of TTC and SMC are identical. 
For the bladder cancer network (bladder), the co-infection network (co-infection) and the MAPK network (MAPK), 
we can only see blue bars because TTC finds smaller control sets than SMC. 
SMC failed to finish the computation for several networks within twelve hours, 
including the network of hematopoietic cell specification (hematopoiesis), the model of mouse embryonic stem cells (PSC-bFA), the CREB network, the model for HGF-induced keratinocyte migration (HGF), the model of bortezomib responses (bortezomib) and the HIV-1 network networks (HIV1). 
For the PC12 cell differentiation network (PC12), the Th-cell differentiation network (T-diff) and the CD4+ T-cell network (CD4+), 
although TTC and SMC require the same number of perturbations, 
our method TTC has the capability to provide more unique solutions, 
which may give more flexibility for practical applications.

\begin{table}[!t]
\centering
\begin{tabular}{L{2.2cm}R{1.8cm}R{1.8cm}R{1.8cm}R{1.8cm}}
\toprule
\multirow{2}{*}{Network} &  \multicolumn{4}{c}{Time (seconds)} \\ \cline{2-5}
 & ITC & TTC &  PTC &	SMC		\\ \midrule
yeast	&	0.028	&	0.987	&	0.933	&	10.837 \\ 
ERBB	&	0.055	&	0.117	&	0.163	&	6.400 \\ 
HSPC-MSC	&	0.097	&	0.101	&	0.109	&	11.393 \\ 
{\small hematopoiesis}	&	0.374	&	139.859	&	72.793	&	$*$ \\ 
PC12	&	0.149	&	17.653	&	22.189	&	234.513 \\ 
bladder	&	0.302	&	2.426	&	7.997	&	36.277 \\ 
psc-bFA	&	36.77	&	3732.78	&	9296.740	&	$*$ \\ 
{\small co-infection}	&	6294.29	&	$*$	&	$*$	&	15097.511 \\ 
MAPK	&	4.608	&	22.218	&	45.504	&	395.014 \\ 
CREB	&	7.962	&	8.277	&	8.693	&	$*$ \\ 
HGF	&	19.925	&	1437.29	&	201.363	&	$*$ \\ 
{\small bortezomib}	&	15.605	&	$*$	&	$*$	&	$*$ \\ 
T-diff	&	21.581	&	29738.5	&	$*$	&	353.473 \\ 
HIV1	&	302.8	&	323.666	&	379.127	&	$*$ \\ 
CD4+	&	549.878	&	1982.45	&	21358.400	&	27.836 \\ 
pathway	&	445.251	&	4435.59	&	10038.600	&	42.180 \\
\bottomrule
\end{tabular}
\caption[Computational time for target control of the biological networks]{Computational time of ITC, TTC, PTC and SMC for several biological networks. Symbol '*' means that the method failed to finish the computation within twelve hours. }
\label{tab:tc2-networks}
\end{table}

\smallskip
\noindent
\textbf{Efficiency.} 
Table~\ref{tab:tc2-networks} summarises the computational time for computing the target control for all the attractors of the networks rather than the selected target attractor. 
The reason is that SMC computes the control for all the attractors in one-go by generating the stable motif diagram, in which different sequences of stable motifs lead to different attractors. 
SMC does not support the computation of target control for only one attractor. 
Hence, for ITC, TTC and PTC, we also take the total computational time for all the attractors of the networks in order to have a fair comparison with SMC.

We can see that ITC is the most efficient one, however, it requires more perturbations. 
TTC and PTC are more efficient than SMC for most of the cases. 
The efficiency of our methods are influenced by many factors, 
such as the network size, the network density, the number of attractors 
and the number of required perturbations. 
For the co-infection network and the model of bortezomib responses, 
TTC and PTC are able to identify target control efficiently for some of the attractors, 
but failed for the other attractors. 
One conjecture is that the target control of those attractors require many perturbations, such that it takes a considerable amount of time to verify the subsets of the schemata.

SMC failed to finish the computation for several networks within twelve hours, 
including the hematopoiesis, PSC-bFA, CREB, HGF, bortezomib and HIV1 networks.  
For the hematopoiesis network and the bortezomib network, 
SMC failed in the identification of stable motifs, 
which has been pointed out to be the most time-consuming part of SMC~\cite{ZA15}. 
The reason could be that the number of cycles and/or SCCs in its expanded network is computationally intractable. 
For the HGF-induced keratinocyte migration network, 
SMC was blocked in the optimisation of stable motifs due to that this network has $19$ stable motifs and
most of the stable motifs contain more than $16$ nodes. 
SMC failed to construct the expanded network representation for the CREB, PSC-BFA and HIV-1 networks 
because some of their Boolean functions depend on many parent nodes ($k \geq 10$). 
Detailed discussion on the complexity of SMC can be found in~\cite{ZA15}.

\section{Conclusion}
\label{sec:conclusion-target-control}

In this paper, we developed three methods for the target control of asynchronous Boolean networks with
instantaneous, temporary and permanent perturbations. 
We compared their performance with 
SMC~\cite{ZA15} on various real-life biological networks. 
The results showed that ITC requires more perturbations than TTC, PTC and SMC as it uses instantaneous perturbations. 
Both TTC and SMC solve the target control problem with temporary perturbations and potentially they may require fewer perturbations than PTC. 
Moreover, compared to SMC, 
our method TTC has the potential to identify more solutions with even fewer perturbations.

Regarding the computational time, 
our methods are quite efficient and scale well for large networks.  
SMC explores both structures and Boolean functions of Boolean networks, and is potentially more scalable for large networks as demonstrated by the CD4+ T-cell network and the pathway network in Table~\ref{tab:tc2-networks}. 
In contrast, our methods are essentially based on the dynamics of the networks,
and they will suffer the state space explosion problem for networks of several hundreds of nodes.
We believe that our methods and SMC complement each other well.
In the near future, we aim to improve our methods
by simultaneously exploring network structure and dynamics
to achieve more efficient computational methods
for the control of large biological networks.
We will also apply our methods for the analysis of real-life biological networks
with the aim of identifying potential drug targets for effective disease treatments.

\section{Acknowledgements}
This work was supported by the project SEC-PBN funded by University of Luxembourg
and the ANR-FNR project AlgoReCell ({\sf INTER/ANR/15/11191283}).



\begin{thebibliography}{10}
\expandafter\ifx\csname url\endcsname\relax
  \def\url#1{\texttt{#1}}\fi
\expandafter\ifx\csname urlprefix\endcsname\relax\def\urlprefix{URL }\fi
\expandafter\ifx\csname href\endcsname\relax
  \def\href#1#2{#2} \def\path#1{#1}\fi

\bibitem{SD16}
D.~Srivastava, N.~DeWitt, In vivo cellular reprogramming: the next generation,
  Cell 166~(6) (2016) 1386--1396.

\bibitem{GD19}
A.~Grath, G.~Dai, Direct cell reprogramming for tissue engineering and
  regenerative medicine, Journal of Biological Engineering 13~(1) (2019) 14.

\bibitem{GMS19}
M.~S. Goligorsky, New trends in regenerative medicine: reprogramming and
  reconditioning, Journal of the American Society of Nephrology 30~(11) (2019)
  2047--2051.

\bibitem{L16}
L.-Z. Wang, R.-Q. Su, Z.-G. Huang, X.~Wang, W.-X. Wang, C.~Grebogi, Y.-C. Lai,
  A geometrical approach to control and controllability of nonlinear dynamical
  networks, Nature Communications 7~(1) (2016) 1--11.

\bibitem{KS69}
S.~Kauffman, Homeostasis and differentiation in random genetic control
  networks, Nature 224 (1969) 177--178.

\bibitem{Aku18}
T.~Akutsu, Algorithms for Analysis, Inference, and Control of Boolean Networks,
  World Scientific, 2018.

\bibitem{HS01}
S.~Huang, Genomics, complexity and drug discovery: insights from {B}oolean
  network models of cellular regulation, Pharmacogenomics 2~(3) (2001)
  203--222.

\bibitem{LSB11}
Y.-Y. Liu, J.-J. Slotine, A.-L. Barab\'asi, Controllability of complex
  networks, Nature 473 (2011) 167--173.

\bibitem{GLDB14}
J.~Gao, Y.-Y. Liu, R.~M. D'Souza, A.-L. Barab\'asi, Target control of complex
  networks, Nature Communications 5 (2014) 5415.

\bibitem{CGCK16}
E.~Czeizler, C.~Gratie, W.~K. Chiu, K.~Kanhaiya, I.~Petre, Target
  controllability of linear networks, in: Proc.\ 14th International Conference
  on Computational Methods in Systems Biology, Vol. 9859 of LNCS, Springer,
  2016, pp. 67--81.

\bibitem{GR16}
A.~J. Gates, L.~M. Rocha, Control of complex networks requires both structure
  and dynamics, Scientific Reports 6~(1) (2016) 1--11.

\bibitem{ABGD13}
A.~Mochizuki, B.~Fiedler, G.~Kurosawa, D.~Saito, Dynamics and control at
  feedback vertex sets. {II}: A faithful monitor to determine the diversity of
  molecular activities in regulatory networks, Journal of Theoretical Biology
  335 (2013) 130--146.

\bibitem{BAGD13}
B.~Fiedler, A.~Mochizuki, G.~Kurosawa, D.~Saito, Dynamics and control at
  feedback vertex sets. {I}: Informative and determining nodes in regulatory
  networks, Journal of Dynamics and Differential Equations 25~(3) (2013)
  563--604.

\bibitem{ZYA17}
J.~G.~T. Za{\~n}udo, G.~Yang, R.~Albert, Structure-based control of complex
  networks with nonlinear dynamics, Proceedings of the National Academy of
  Sciences 114~(28) (2017) 7234--7239.

\bibitem{CKM13}
S.~P. Cornelius, W.~L. Kath, A.~E. Motter, Realistic control of network
  dynamics, Nature Communications 4~(1) (2013) 1--9.

\bibitem{LCL17}
J.~Liang, H.~Chen, J.~Lam, An improved criterion for controllability of
  {B}oolean control networks, IEEE Transactions on Automatic Control 62~(11)
  (2017) 6012--6018.

\bibitem{ZLLC18}
Q.~Zhu, Y.~Liu, J.~Lu, J.~Cao, Further results on the controllability of
  {B}oolean control networks, IEEE Transactions on Automatic Control 64~(1)
  (2018) 440--442.

\bibitem{LZHY16}
J.~Lu, J.~Zhong, D.~W. Ho, Y.~Tang, J.~Cao, On controllability of delayed
  {B}oolean control networks, SIAM Journal on Control and Optimization 54~(2)
  (2016) 475--494.

\bibitem{ZLKS19}
J.~Zhong, Y.~Liu, K.~I. Kou, L.~Sun, J.~Cao, On the ensemble controllability of
  {B}oolean control networks using {STP} method, Applied Mathematics and
  Computation 358 (2019) 51--62.

\bibitem{WSZS19}
Y.~Wu, X.-M. Sun, X.~Zhao, T.~Shen, Optimal control of {B}oolean control
  networks with average cost: A policy iteration approach, Automatica 100
  (2019) 378--387.

\bibitem{CLW16}
H.~Chen, J.~Liang, Z.~Wang, Pinning controllability of autonomous {B}oolean
  control networks, Science China Information Sciences 59~(7) (2016) 070107.

\bibitem{YYCJ19}
J.~Yue, Y.~Yan, Z.~Chen, X.~Jin, Identification of predictors of {B}oolean
  networks from observed attractor states, Mathematical Methods in the Applied
  Sciences 42~(11) (2019) 3848--3864.

\bibitem{ZKF13}
Y.~Zhao, J.~Kim, M.~Filippone, Aggregation algorithm towards large-scale
  {B}oolean network analysis, IEEE Transactions on Automatic Control 58~(8)
  (2013) 1976--1985.

\bibitem{KSK13}
J.~Kim, S.-M. Park, K.-H. Cho, Discovery of a kernel for controlling
  biomolecular regulatory networks, Scientific Reports 3 (2013) 2223.

\bibitem{MGF19}
M.~Moradi, S.~Goliaei, M.-H. Foroughmand-Araabi, A {B}oolean network control
  algorithm guided by forward dynamic programming, PLOS ONE 14~(5) (2019)
  e0215449.

\bibitem{PSPM18}
S.~Paul, C.~Su, J.~Pang, A.~Mizera, A decomposition-based approach towards the
  control of {B}oolean networks, in: Proc.\ 9th {ACM} {C}onference on
  {B}ioinformatics, {C}omputational {B}iology, and {H}ealth {I}nformatics, ACM
  Press, 2018, pp. 11--20.

\bibitem{PSPM19}
S.~Paul, C.~Su, J.~Pang, A.~Mizera, An efficient approach towards the
  source-target control of {B}oolean networks, IEEE/ACM Transactions on
  Computational Biology and Bioinformatics 17~(6) (2020) 1932--1945.

\bibitem{SPP19b}
C.~Su, S.~Paul, J.~Pang, Controlling large {B}oolean networks with temporary
  and permanent perturbations, in: Proc.\ 23rd International Symposium on
  Formal Methods, Vol. 11800 of LNCS, Springer, 2019, pp. 707--724.

\bibitem{MSPPHP19}
H.~Mandon, C.~Su, J.~Pang, S.~Paul, S.~Haar, L.~Paulev\'{e}, Algorithms for the
  sequential reprogramming of {B}oolean networks, IEEE/ACM Transactions on
  Computational Biology and Bioinformatics 16~(5) (2019) 1610--1619.

\bibitem{MSHPP19}
H.~Mandon, C.~Su, S.~Haar, J.~Pang, L.~Paulev\'e, Sequential reprogramming of
  {B}oolean networks made practical, in: Proc.\ 17th {I}nternational
  {C}onference on {C}omputational {M}ethods in {S}ystems {B}iology, Vol. 11773
  of LNCS, Springer, 2019, pp. 3--19.

\bibitem{SC14}
A.~del Sol, N.~J. Buckley, Concise review: A population shift view of cellular
  reprogramming, Stem Cells 32~(6) (2014) 1367--1372.

\bibitem{BCB20}
C.~Su, J.~Pang, A dynamics-based approach for the target control of {B}oolean
  networks., in: Proc.\ 11th {ACM} {C}onference on {B}ioinformatics,
  {C}omputational {B}iology, and {H}ealth {I}nformatics, ACM Press, 2020, pp.
  50:1--50:8.

\bibitem{ZA15}
J.~G. Za\~nudo, R.~Albert, Cell fate reprogramming by control of intracellular
  network dynamics, PLOS Computational Biology 11~(4) (2015) e1004193.

\bibitem{MS11}
F.-J. M{\"u}ller, A.~Schuppert, Few inputs can reprogram biological networks,
  Nature 478~(7369) (2011) E4.

\bibitem{MPSY17}
A.~Mizera, J.~Pang, C.~Su, Q.~Yuan, {ASSA-PBN}: A toolbox for probabilistic
  {B}oolean networks, IEEE/ACM Transactions on Computational Biology and
  Bioinformatics 15~(4) (2018) 1203--1216.

\bibitem{ZH14}
P.~Zhu, J.~Han, Asynchronous stochastic {B}oolean networks as gene network
  models, Journal of Computational Biology 21~(10) (2014) 771--783.

\bibitem{SP2020}
C.~Su, J.~Pang, Sequential temporary and permanent control of {B}oolean
  networks., in: Proc.\ 18th {I}nternational {C}onference on {C}omputational
  {M}ethods in {S}ystems {B}iology, Vol. 12314 of LNCS, Springer, 2020, pp.
  234--251.

\bibitem{cudd}
F.~Somenzi, {CUDD}: {CU} decision diagram package - release 2.5.1,
  http://vlsi.colorado.edu/~fabio/CUDD/ (2015).

\bibitem{cabean}
C.~Su, J.~Pang, {CABEAN}: a software for the control of asynchronous {B}oolean
  networks, Bioinformatics(accepted).

\bibitem{MPQY17b}
A.~Mizera, J.~Pang, H.~Qu, Q.~Yuan, Taming asynchrony for attractor detection
  in large {B}oolean networks, IEEE/ACM Transactions on Computational Biology
  and Bioinformatics 16~(1) (2019) 31--42.

\bibitem{YMPQ19}
Q.~Yuan, A.~Mizera, J.~Pang, H.~Qu, A new decomposition-based method for
  detecting attractors in synchronous boolean networks, Science of Computer
  Programming 180 (2019) 18--35.

\bibitem{assa}
A.~Mizera, J.~Pang, Q.~Yuan, {ASSA-PBN}: a tool for approximate steady-state
  analysis of large probabilistic {B}oolean networks, in: Proc.\ 13th
  International Symposium on Automated Technology for Verification and
  Analysis, Vol. 9364 of LNCS, Springer, 2015, pp. 214--220.

\bibitem{MPQY18}
A.~Mizera, J.~Pang, H.~Qu, Q.~Yuan, {ASSA-PBN} 3.0: {A}nalysing
  context-sensitive probabilistic {B}oolean networks, in: Proc.\ 16th
  International Conference on Computational Methods in Systems Biology, Vol.
  11095 of LNCS, Springer, 2018, pp. 313--317.

\bibitem{KMST11}
J.~Krumsiek, C.~Marr, T.~Schroeder, F.~J. Theis, Hierarchical differentiation
  of myeloid progenitors is encoded in the transcription factor network, PLOS
  ONE 6~(8) (2011) e22649.

\bibitem{HGZKK12}
F.~Herrmann, A.~Gro{\ss}, D.~Zhou, H.~A. Kestler, M.~K{\"u}hl, A {Boolean}
  model of the cardiac gene regulatory network determining first and second
  heart field identity, PLOS ONE 7~(10) (2012) e46798.

\bibitem{DB08}
M.~I. Davidich, S.~Bornholdt, Boolean network model predicts cell cycle
  sequence of fission yeast, PLOS ONE 3~(2) (2008) e1672.

\bibitem{SFLK09}
{\"O}.~Sahin, H.~Fr{\"o}hlich, C.~L{\"o}bke, U.~Korf, S.~Burmester, M.~Majety,
  J.~Mattern, I.~Schupp, C.~Chaouiya, D.~Thieffry, A.~Poustka, S.~Wiemann,
  T.~Beissbarth, D.~Arlt, Modeling {ERBB} receptor-regulated {G}1/{S}
  transition to find novel targets for de novo trastuzumab resistance, BMC
  Systems Biology 3~(1) (2009) 1.

\bibitem{EMMP16}
J.~Enciso, H.~Mayani, L.~Mendoza, R.~Pelayo, Modeling the pro-inflammatory
  tumor microenvironment in acute lymphoblastic leukemia predicts a breakdown
  of hematopoietic-mesenchymal communication networks, Frontiers in Physiology
  7 (2016) 349.

\bibitem{CMC15}
D.~P. Cohen, L.~Martignetti, S.~Robine, E.~Barillot, A.~Zinovyev, L.~Calzone,
  Mathematical modelling of molecular pathways enabling tumour cell invasion
  and migration, PLOS Computational Biology 11~(11) (2015) e1004571.

\bibitem{COOAD17}
S.~Collombet, C.~van Oevelen, J.~L.~S. Ortega, W.~Abou-Jaoud{\'e},
  B.~Di~Stefano, M.~Thomas-Chollier, T.~Graf, D.~Thieffry, Logical modeling of
  lymphoid and myeloid cell specification and transdifferentiation, Proceedings
  of the National Academy of Sciences 114~(23) (2017) 5792--5799.

\bibitem{OKB16}
B.~Offermann, S.~Knauer, A.~Singh, M.~L. Fern{\'a}ndez-Cach{\'o}n, M.~Klose,
  S.~Kowar, H.~Busch, M.~Boerries, {Boolean} modeling reveals the necessity of
  transcriptional regulation for bistability in {PC12} cell differentiation,
  Frontiers in Genetics 7 (2016) 44.

\bibitem{WSA12}
E.~Remy, S.~Rebouissou, C.~Chaouiya, A.~Zinovyev, F.~Radvanyi, L.~Calzone, A
  modeling approach to explain mutually exclusive and co-occurring genetic
  alterations in bladder tumorigenesis, Cancer Research 75~(19) (2015)
  4042--4052.

\bibitem{YOOLP18}
A.~Yachie-Kinoshita, K.~Onishi, J.~Ostblom, M.~A. Langley, E.~Posfai,
  J.~Rossant, P.~W. Zandstra, Modeling signaling-dependent pluripotency with
  boolean logic to predict cell fate transitions, Molecular Systems Biology
  14~(1) (2018) e7952.

\bibitem{TPMA12}
J.~Thakar, A.~K. Pathak, L.~Murphy, R.~Albert, I.~M. Cattadori, Network model
  of immune responses reveals key effectors to single and co-infection dynamics
  by a respiratory bacterium and a gastrointestinal helminth, PLOS
  Computational Biology 8~(1) (2012) e1002345.

\bibitem{GCT13}
L.~Grieco, L.~Calzone, I.~Bernard-Pierrot, F.~Radvanyi, B.~Kahn-Perles,
  D.~Thieffry, Integrative modelling of the influence of {MAPK} network on
  cancer cell fate decision, PLOS Computational Biology 9~(10) (2013) e1003286.

\bibitem{ATE08}
A.~Abdi, M.~B. Tahoori, E.~S. Emamian, Fault diagnosis engineering of digital
  circuits can identify vulnerable molecules in complex cellular pathways,
  Science Signaling 1~(42) (2008) ra10--ra10.

\bibitem{SNB12}
A.~Singh, J.~M. Nascimento, S.~Kowar, H.~Busch, M.~Boerries, {Boolean} approach
  to signalling pathway modelling in {HGF-}induced keratinocyte migration,
  Bioinformatics 28~(18) (2012) 495--501.

\bibitem{COAM15}
V.~L. Chudasama, M.~A. Ovacik, D.~R. Abernethy, D.~E. Mager, Logic-based and
  cellular pharmacodynamic modeling of bortezomib responses in {U}266 human
  myeloma cells, Journal of Pharmacology and Experimental Therapeutics 354~(3)
  (2015) 448--458.

\bibitem{NCCT10}
A.~Naldi, J.~Carneiro, C.~Chaouiya, D.~Thieffry, Diversity and plasticity of th
  cell types predicted from regulatory network modelling, PLOS Computational
  Biology 6~(9) (2010) e1000912.

\bibitem{ODRS14}
O.~J. Oyeyemi, O.~Davies, D.~L. Robertson, J.-M. Schwartz, A logical model of
  {HIV}-1 interactions with the {T}-cell activation signalling pathway,
  Bioinformatics 31~(7) (2014) 1075--1083.

\bibitem{conroy2014design}
B.~D. Conroy, T.~A. Herek, T.~D. Shew, M.~Latner, J.~J. Larson, L.~Allen, P.~H.
  Davis, T.~Helikar, C.~E. Cutucache, Design, assessment, and in vivo
  evaluation of a computational model illustrating the role of {CAV1} in {CD4+}
  {T}-lymphocytes, Frontiers in Immunology 5 (2014) 599.

\bibitem{RRLP08}
S.~Raza, K.~A. Robertson, P.~A. Lacaze, D.~Page, A.~J. Enright, P.~Ghazal,
  T.~C. Freeman, A logic-based diagram of signalling pathways central to
  macrophage activation, BMC Systems Biology 2~(1) (2008) 36.

\end{thebibliography}

\end{document}